\documentclass[twocolumn]{aastex63}
\usepackage{amsopn} 
\usepackage{comment}

\usepackage[normalem]{ulem}

\usepackage{amsmath}

\newcommand\guidoni{GUID16}
\DeclareMathOperator\erf{erf}

\newcommand{\tablespaceadd}{\rule{0pt}{6ex}}  
\newcommand{\tablespaceaddtwo}{\rule{0pt}{3ex}}

\shorttitle{Spectral Power-law Formation by Sequential Particle Acceleration}
\shortauthors{Guidoni et al.}

\begin{document}

\title{Spectral Power-law Formation by Sequential Particle Acceleration in Multiple Flare Magnetic Islands}

\correspondingauthor{Silvina E. Guidoni}
\email{guidoni@american.edu}

\author[0000-0003-1439-4218]{S.\ E.\ Guidoni}
\affil{Department of Physics \\ American University \\
4400 Massachusetts Avenue NW\\
Washington, DC 20016, USA}
\affiliation{Heliophysics Science Division\\
NASA Goddard Space Flight Center \\ 8800 Greenbelt Road \\
Greenbelt, MD 20771, USA}

\author{J.\ T.\ Karpen}
\affiliation{Heliophysics Science Division\\
NASA Goddard Space Flight Center \\ 8800 Greenbelt Rd \\
Greenbelt, MD 20771, USA}

\author{C.\ R.\ DeVore }
\affiliation{Heliophysics Science Division\\
NASA Goddard Space Flight Center \\ 8800 Greenbelt Rd \\
Greenbelt, MD 20771, USA}

\begin{abstract}

We present a first-principles model of pitch-angle and energy distribution function evolution as particles are sequentially accelerated by multiple flare magnetic islands. Data from magnetohydrodynamic (MHD) simulations of an eruptive flare/coronal mass ejection provide ambient conditions for the evolving particle distributions. Magnetic islands, which are created by sporadic reconnection at the self-consistently formed flare current sheet, contract and accelerate the particles. The particle distributions are evolved using rules derived in our previous work. In this investigation, we assume that a prescribed fraction of particles sequentially ``hops'' to another accelerator and receives an additional boost in energy and anisotropy. This sequential process generates particle number spectra that obey an approximate power law at mid-range energies and presents low- and high-energy breaks. We analyze these spectral regions as functions of the model parameters. We also present a fully analytic method for forming and interpreting such spectra, \textit{independent of the sequential acceleration model}. The method requires only a few constrained physical parameters, such as the percentage of particles transferred between accelerators, the energy gain in each accelerator, and the number of accelerators visited. Our investigation seeks to bridge the gap between MHD and kinetic regimes by combining global simulations and analytic kinetic theory. The model reproduces and explains key characteristics of observed flare hard X-ray spectra as well as the underlying properties of the accelerated particles. Our analytic model provides tools to interpret high-energy observations for missions and telescopes, such as RHESSI, FOXSI, NuSTAR, Solar Orbiter, EOVSA, and future high-energy missions.  
\end{abstract}


\section{Introduction}
\label{sec:intro}

Sudden large-scale reconfigurations of the solar coronal magnetic field manifest as the most powerful explosions in the solar system: eruptive solar flares (EFs) and coronal mass ejections (CMEs). Flare emissions are observed across the electromagnetic spectrum, from $\gamma$ rays to radio waves. Understanding the mechanism that efficiently accelerates prodigious numbers of electrons to the high energies required to produce the observed flare $\gamma$-ray, hard X-ray (HXR), and microwave emissions is a long-sought goal in heliophysics. Observations point indirectly to magnetic reconnection as the fundamental process involved in flare particle acceleration \citep[see review by][]{Zharkova_2011}, but the  mechanism that transfers the released magnetic energy to ambient electrons and ions remains under debate. 

In the standard flare model \citep{Carmichael_1964, Sturrock_1966, Hirayama_1974, Kopp_1976}, oppositely directed field lines reconnect across a large-scale current sheet. Particles could be accelerated directly by the current-sheet electric field, in the flows driven by the retracting field lines, by shocks, or by the merger or contraction of islands formed by reconnection in the current sheet. The work presented here is focused on the last mechanism. 

Flare X-rays are emitted predominantly by high-energy electrons scattering off background ions (bremsstrahlung). The source electrons are generally agreed to be energized in the corona, but most of the observed HXR radiation emanates from flare arcade footpoints where the accelerated particles encounter the dense chromosphere and photosphere. This is the so-called thick-target model for X-ray production \citep{Brown_1971}. When this dominant source is occulted, however, HXR emission is also observed above the top of the soft X-ray loops \citep[e.g.,][]{Masuda_1994, Krucker_2010}, both below and above the presumed reconnection site \citep[e.g.,][]{Battaglia_2019}. 

Typically, the flare X-ray energy spectrum can be divided into two components: 1) at low energies, a thermal component emitted by bulk flare-heated plasma; and 2) at higher energies, a non-thermal power-law component (or double power law, \citet{Alaoui_2019}), $\epsilon^{-\gamma}$, where $\epsilon$ is the photon energy and $\gamma$ is the photon spectral index. The index usually falls in the range $\gamma \sim$ 2-10 \citep{Brown_1971, Dennis_1985, Petrosian_2002, Holman_2003, Krucker_2008, Krucker_2008_rev, Hannah_2008, Christe_2008}. 

The differential energy of the electrons responsible for the nonthermal portion of the HXR spectrum is generally assumed to follow a power law, $E^{-\delta'}$ \citep{Holman_2003_I}, where $E$ is the electron energy and $\delta'$ is the spectral index  (to avoid confusion, we are using the notation of \citet{Oka_2018} for spectral indices). To ensure that the energy of the injected electrons is finite, the electron spectrum is usually assumed to cut off sharply or flatten below a low-energy cutoff \citep{Holman_2003_I,  Kontar_2008, Alaoui_2017, McTiernan_2019}. Some observations also indicate the need for a cutoff or other change in the spectral shape at high energies \citep[e.g.,][]{Holman_2003_I}. The total energy in the accelerated electrons strongly depends on the cutoff energies and on the shape of the distribution at low energies \citep{Emslie_2003, Saint_Hilaire_2005, Galloway_2005}. The relationship between the photon and electron energy spectral indices depends on how particles lose their energy as they interact with the ambient plasma. A thick-target source yields $\gamma_{thick} = \delta'-3/2$ \citep{Brown_1971,Hudson_1972}, whereas for a thin-target source $\gamma_{thin} = \delta'+1/2$ \citep{Tandberg_Hanssen_1988}. Recent advances in particle-ambient interactions have taken into account propagation mechanisms such as return-current losses \citep{Alaoui_2017}, energy diffusion in a ``warm'' target \citep{Kontar_2015}, and nonuniform ionization of the thick target \citep{Su_2011}.

Observations of rapid temporal intermittency in HXR and microwaves during the flare's impulsive phase \citep{Inglis_2012, Inglis_2013, Inglis_2016, Hayes_2016, Hayes_2019}, as well as bright plasma blobs traveling in both directions along the flare current sheet, provide strong evidence for the formation of magnetic islands during flare reconnection and particle acceleration within them \citep{Kliem_2000, Karlicky_2004, Karlicky_2007, Barta_2008,  Liu_2013, Kumar_2013, Takasao_2016,  Kumar_2013_I, Zhao_2019}. Numerous theoretical and high-resolution numerical studies have demonstrated that extended current sheets with large Lundquist numbers develop multiple reconnection sites with strong spatial and temporal variability on both kinetic and magnetohydrodynamic (MHD) scales \citep[e.g.,][]{Daughton_2006, Daughton_2014, Drake_2006_I, Loureiro_2007, Samtaney_2009, Fermo_2010, Uzdensky_2010, Huang_2012, Mei_2012, Cassak_2013, Shen_2013}. 
 
Kinetic-scale particle-in-cell (PIC) simulations have shown that particles can be energized in contracting and merging magnetic islands \citep{Drake_2005,Drake_2006, Drake_2006_I, Drake_2010, Drake_2013,Dahlin_2016, Dahlin_2017}, and that the resulting electron energy spectra can achieve power laws \citep{Guo_2015, Ball_2018, Li_2019}. However, even the most advanced PIC simulations \citep{Daughton_2014, Guo_2015} are incapable of modeling the large dimensions and numbers of particles involved in flares \citep{Dahlin_2017}. 

In \citet{Guidoni_2016} (henceforth referred to as \guidoni), we applied the contracting-island scenario to a simulated eruptive solar flare where intermittent reconnection forms macroscopic islands \citep{Karpen_2012}. Combining analytical calculations for individual test particles with data from the global simulation, which self-consistently modeled formation and reconnection onset at the flare current sheet, we found that compression and contraction of a single island increased the particle energies by a factor up to $\sim 5$. The results were confirmed subsequently by numerically integrating the particle guiding-center trajectories \citep{Borovikov_2017}. 
Although these initial findings were encouraging, such small energy boosts are insufficient to produce either the required energies or power laws needed to explain flare emission spectra. 

The objective of this paper is to construct and evolve distribution functions as particles are accelerated sequentially by several magnetic islands in the flare current sheet. The ambient particle distribution is assumed to be Maxwellian initially. It evolves as particles ``hop'' from one contracting island to another, receiving a moderate energy boost in each island. We demonstrate analytically that this mechanism can generate power-law indices, high-energy cutoffs, and flat low-energy spectra consistent with observations of solar flares. 

\section{Particle Acceleration in a Single Magnetic Island} 
\label{sec:previous}

Here we briefly describe the relevant results from \guidoni~ and add figures, calculations, and explanations needed for the present work. In that study, we developed an analytic method to estimate energy gain for particles assumed to be orbiting within single flux ropes formed by flare magnetic reconnection in an MHD simulation of a breakout solar eruption. The method is based on the assumption that the particles' parallel action and magnetic moment are conserved as particles gyrate around magnetic field lines, and is applicable to moderately superthermal electrons and strongly superthermal ions.

The evolving flux-rope properties were extracted from an ultra-high-resolution (8 levels of refinement), cylindrically axisymmetric (2.5D), global MHD numerical simulation of a CME/EF using the Adaptively Refined MHD Solver \citep[ARMS; e.g.,][]{DeVore_2008}. According to the well-established breakout CME model \citep{Antiochos_1998, Antiochos_1999_I}, a multipolar active-region magnetic field forms a filament channel by  shearing (through motions or helicity condensation) of the field immediately surrounding the polarity inversion line. The stressed core flux expands and distorts the overlying null into a current sheet, enabling breakout reconnection that removes restraints on the rising core. As the filament-channel flux stretches out into the corona, a lengthening flare current sheet (CS) forms beneath it, leading to flare reconnection.  Field lines retracting sunward after the onset of fast flare reconnection create the flare arcade, while those retracting in the opposite direction form the large CME flux rope \citep[for more details see][and GUID16]{Karpen_2012}.

Temporally and spatially intermittent reconnection across the flare CS forms small flux ropes (islands, in 2.5D), which are expelled along the CS in opposite directions from a slowly rising main reconnection null. 

We found little evidence for island merging, in contrast to kinetic simulations of reconnection in pre-existing current sheets with periodic boundary conditions \citep[e.g.,][]{Drake_2006}. 

\begin{figure}[ht]
\centering
\includegraphics[width=\columnwidth]{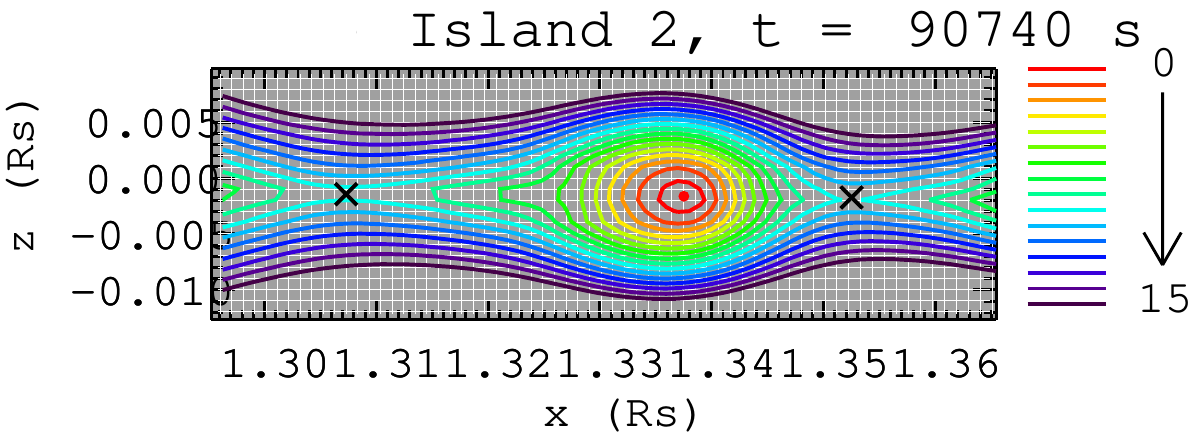}
   \caption{Snapshot of Island 2. Selected flux surfaces (accelerators) are color-coded and labeled from 0 to 15 (as shown at right) from innermost to outermost with respect to the island's O-null (red dot). Black crosses are X-null locations. The simulation grid is shown in white. The x-axis is parallel to the plane of the flare CS, and the z-axis perpendicular to the CS plane. Both axes are in units of solar radius $R_S$. Flux surface level ``8'' is referred to as accelerator A2 in this paper. }
   \label{fig:Island_2}
\end{figure}

\begin{figure*}[ht]
\centering
\includegraphics[width=\textwidth]{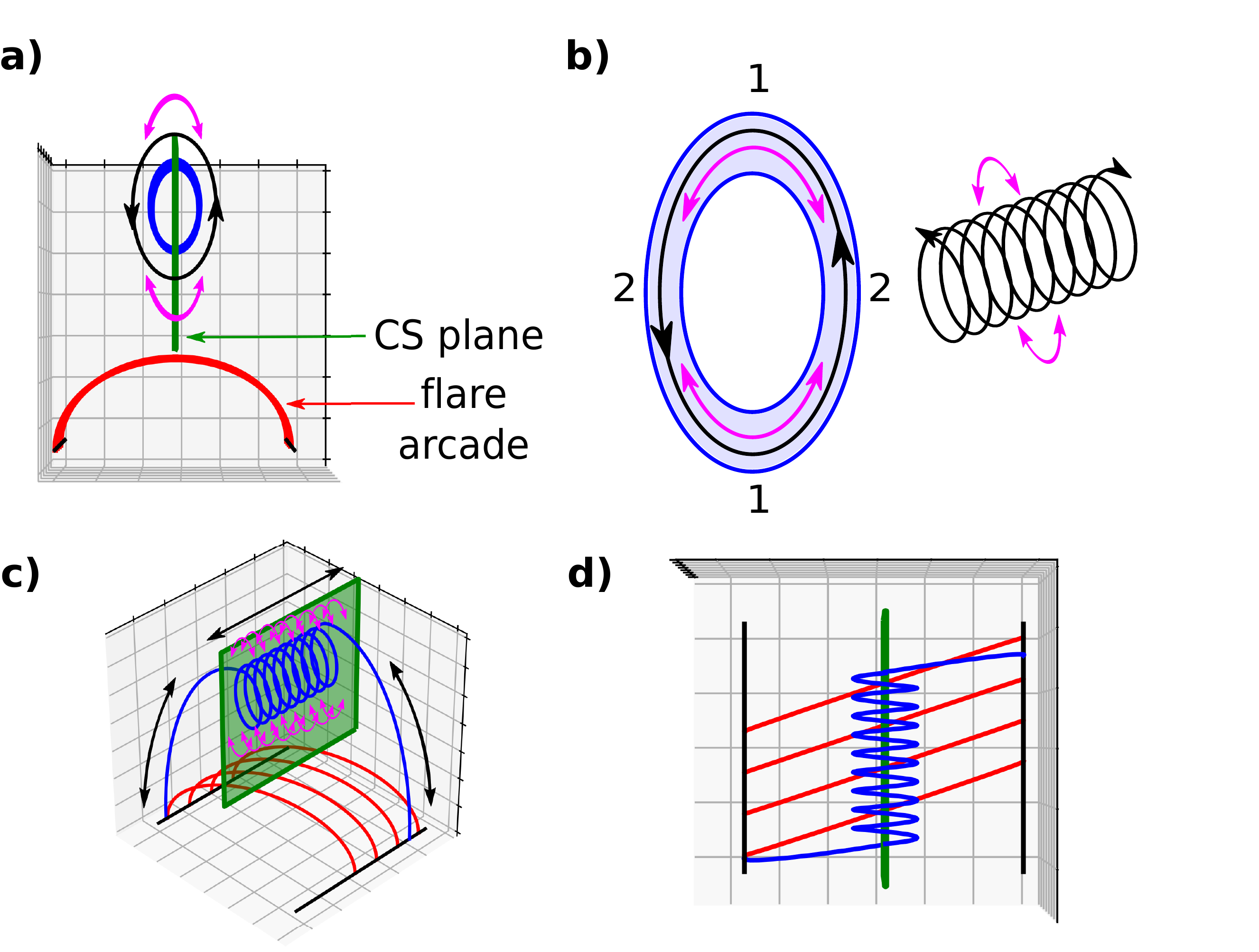}
   \caption{2.5D and 3D illustrations of the flare system, its accelerators, and particle trajectories. In all panels, a representative flux-rope field line is shown in blue, the flare CS plane in green, and the sheared flare arcade in red. Transiting (mirroring) particle trajectories are shown with curved black (magenta) arrows. The kinetic-scale gyration of the particles around the field lines is not shown. a) Key features of a 2.5D system projected onto a plane perpendicular to the flare CS plane (translationally invariant direction is out of the image plane). Here the flare CS is vertical, whereas the flare CS is horizontal in Figure \ref{fig:Island_2}. b) Left: expanded view of the cross section (light blue) of an accelerator. Numbered locations are explained in the text. Right: angled view of 2.5D trajectories of mirroring (one field-line turn) and transiting particles. c) Angled view of a 3D island: a flux rope with a finite axial length. Black arrows show the overall direction of motion for transiting particles along the flux rope. d) Top view of c).}
   \label{fig:cartoon_both_populations}
\end{figure*}

We studied two long-lived, well-resolved, sunward-moving islands, named ``Island 1'' and ``Island 2''. Figure \ref{fig:Island_2} shows color-coded flux surfaces of Island 2 at a time between its formation and its arrival at the top of the flare arcade. The $z=0$ plane corresponds to the plane of the flare CS. 
The island's enclosed magnetic flux, delimited by the flux surfaces near the X-nulls (black Xs in Figure \ref{fig:Island_2}), was elongated along the CS (note horizontal and vertical scale differences in Figure \ref{fig:Island_2}). The islands evolved to a rounder configuration due to the Lorentz force acting on the highly bent field lines near the tapered ends. The field lines on each side of the CS confine an island and limit its expansion perpendicular to the CS. The island's cross-sectional area shrinks, thereby increasing the magnetic field strength. As a result, particles orbiting the island are accelerated mostly by the betatron process, which relies on magnetic-field compression rather than Fermi acceleration \citep[GUID16;][]{Borovikov_2017,Li_2018,Li_2019}.

When scaled to average active-region sizes and characteristic times, the lifetime of these simulated islands, defined as the time between their creation by two adjacent reconnection episodes and their arrival at the top of the flare arcade, is of the order of 10-15 s. Similarly, their typical lengths along the flare CS (x-axis in Figure \ref{fig:Island_2}) are $\sim$ (2-4)$\times 10^{-3}R_s \approx$ 2-4 arcsec, where $R_s$ is the solar radius. Flare plasmoids of similar sizes  have been observed \citep{Kumar_2013}, and typical HXR pulsation periods are comparable to these island lifetimes \citep{Inglis_2012, Inglis_2013, Inglis_2016, Hayes_2016, Hayes_2019},. 

Particles were assumed to be frozen-in, orbiting field lines wrapping selected flux surfaces of the studied islands. Each flux surface represents a finite volume of plasma inside a cylinder-like shell of small thickness, which we denote as ``accelerator''. Figure \ref{fig:cartoon_both_populations}a illustrates a generic field line of such a flux rope (blue) as the flare CS (green) above the sheared flare arcade (red) is viewed head-on.
Figure \ref{fig:cartoon_both_populations}b illustrates the cross-sectional area (light blue) of a generic accelerator. 

In \guidoni, we parameterized the selected accelerators' magnetic-field strength along representative flux-rope field lines as
\begin{equation}
   B = B_{1} + \left( B_{2}-B_{1} \right) \sin^{2} \left( 2 \pi \frac{l}{L} \right),
    \label{eq:B_prof}
\end{equation}
where $L$ is the length of one full turn of the field line and $l$ is the field line arc-length. The flux surface is symmetric both left/right and up/down and possesses two equal minima in $B$ near the X-nulls (labeled ``1'' in Figure \ref{fig:cartoon_both_populations}b) and two equal maxima in $B$ at its points furthest from the CS plane (labeled ``2'' in the same figure). $B_{1}$ and $B_{2}$ are the minimum and maximum field strengths, respectively, at those locations.  

We extracted the evolution of $L$, $B_{1}$, and $B_{2}$ over each accelerator's lifetime from the simulation data. $L$ decreased rapidly as the flux surface contracted and $B_{1}$ increased due to plasma compression. The evolution of $B_{2}$ resulted in an accelerator mirror ratio, $B_{2}/B_{1}$, that was initially larger than $1$ ($\leq 1.6$ for Island 1 and $\leq 8$ for Island 2) and decreased to values close to $1$ as the islands became circular.

Two distinct particle populations orbit each accelerator: \textit{transiting} and \textit{mirroring}. If a particle's pitch angle is smaller (larger) than the loss-cone angle of the accelerator, defined as $\theta^{\hbox{lc}} = \arcsin\left({\sqrt{B_{1}/B_{2}}}\right)$, the particle transits (mirrors) along the accelerator. Mirroring particles bounce at regions of high field strength; their trajectories are marked by the curved magenta arrows in Figures \ref{fig:cartoon_both_populations}a,b. For visual simplicity, particle gyromotions are not shown. As long as the mirror ratio is larger than unity, relatively large populations of mirroring particles can be trapped near the plane of the flare CS. The length of the flux-rope axis does not matter in this case because mirroring particles are trapped in \textit{a single turn} of the flux rope (see example on the right side of Fig.\ \ref{fig:cartoon_both_populations}b.) As the islands are carried by the reconnection exhaust, mirroring particles stay near the flare CS region until the island merges with the top of the flare arcade or the bottom of the CME.

Transiting particles follow the helical field lines, as illustrated by curved black arrows in Figures \ref{fig:cartoon_both_populations}a,b. In a translationally invariant (2.5D) simulation (e.g., \guidoni), transiting particles are trapped in the toroidal flux rope. In a 3D configuration, where the flux rope is anchored at the solar surface, transiting particles are free to stream along the legs of the flux rope and could be lost at the footpoints before they are accelerated. Some of this streaming population could mirror near the footpoints due to the increase in field strength with decreasing altitude (not considered here or in \guidoni). Figures \ref{fig:cartoon_both_populations}c,d illustrate lateral and top views of a flux rope with a finite length axis (3D island) and the overall trajectories of transiting (black) and mirroring (magenta) populations.  

As particles orbit the time-dependent field line described by Eq.\ \ref{eq:B_prof}, their kinetic energy $E$ and pitch angle $\theta$ change. Assuming conservation of the particle parallel action and magnetic moment, \guidoni~estimated the changes in $E$ and $\theta$ as particles pass location ``1'' of the accelerator. Henceforth, all initial and final kinetic energies and pitch angles refer to this location. Only pitch angles $0 \leq \theta \leq 90^\circ$ were considered as the system is symmetric about $\theta = 90^\circ$ (parallel or antiparallel motion with respect to the magnetic field). 

\begin{figure}[ht]
\centering
\includegraphics[width=\columnwidth]{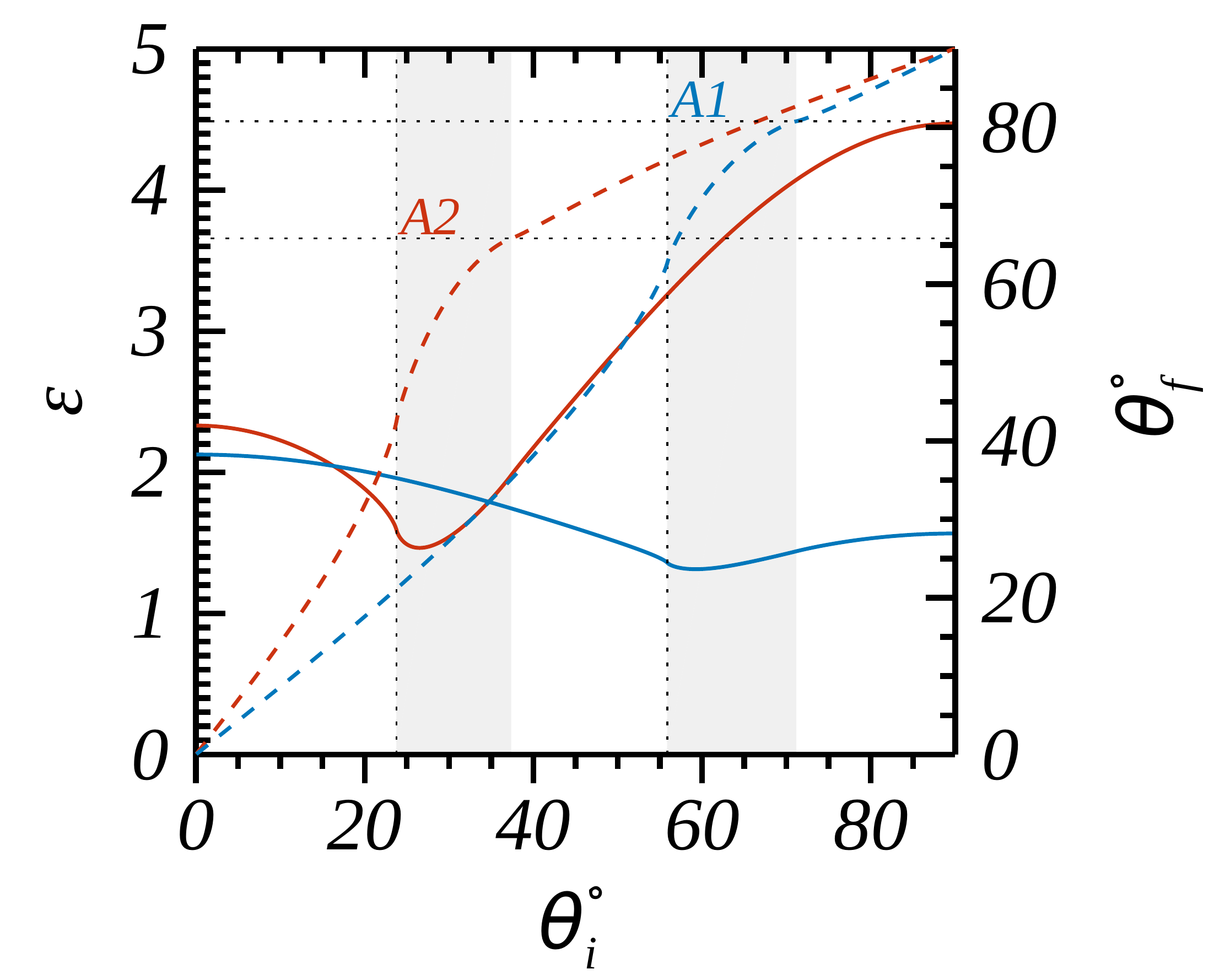}
   \caption{Energy ratios (solid lines, left axis) and final pitch angle (dashed lines, right axis) as functions of $\theta_{i}$ for accelerators A1 (blue) and A2 (red). Final values correspond to the end of the lifetime of both accelerators. Vertical (horizontal) dotted black lines show initial (final) $\theta^{\hbox{lc}}$ for the accelerator labeled where the dotted lines intersect. Light gray areas show ranges in pitch angle where initially mirroring particles are transiting at the final time. Files named ``A1\_energy\_gain\_and\_pitch\_angle\_data.txt'' and ``A2\_energy\_gain\_and\_pitch\_angle\_data.txt'' with the data of this figure are part of the supplemental material of this publication. Results presented in Sections \ref{sec:PALMI} and \ref{sec:results} are based on these data.}
   \label{fig:energization_rules}
\end{figure}

We determined the final pitch angle $\theta_{f}$ and final-to-initial energy ratio $\mathcal{E} = E_{f}/E_{i}$ as functions of the initial pitch angle, $\theta_{i}$, by solving Equations 25 and 26 in \guidoni.  These transcendental equations depend on $L$, $B_{1}$, and $B_{2}$, which we obtained from the simulation. Examples of $\mathcal{E}$ and $\theta_{f}$ as functions of $\theta_{i}$ are shown as solid and dashed curves, respectively, in Figure \ref{fig:energization_rules}. Results presented in Sections \ref{sec:PALMI} and \ref{sec:results} are based on these data. The blue (red) lines represent the selected accelerator in Island 1 (2) labeled ``1'' (``8'')  in \guidoni, which we refer to here as A1 (A2). A2 corresponds to the outermost green flux surface inside the island of Figure \ref{fig:Island_2}. An initially isotropic distribution in pitch angle would be anisotropic at the end of the lifetime of both accelerators (in Figure \ref{fig:energization_rules}, dashed curves differ from straight lines of slope $1$), as shown in the next Section. 

A1's (A2's) initial and final $\theta^{\hbox{lc}}$ are $\simeq 56^\circ$ ($\simeq 24^\circ$) and $\simeq 80^\circ$ ($\simeq 66^\circ$), respectively (shown with dotted lines in Fig.\ \ref{fig:energization_rules}). For initially isotropic distributions, $\simeq 38\%$ ($\simeq 74\%$) of A1's (A2's) population would be mirroring. The maximum energy gain overall for A1 (A2) is $\mathcal{E}_{max} \simeq 2.13$ ($\simeq 4.47$). For mirroring populations, A1's (A2's) maximum energy gain occurs at $90^\circ$ with $\mathcal{E}_{max} \simeq 1.57$ ($\simeq 4.47$). For all of the studied cases in \guidoni, $\mathcal{E}$ varied from one flux surface to another, reaching a maximum value $\mathcal{E}_{max} < 5$. 

As pointed out in \guidoni, such small energy gains are well below the magnitudes required to explain the observed flux and power-law index of flare electron energy spectra. However, particles may increase their energy substantially by ``visiting'' a few accelerators sequentially. For example, visiting only five accelerators with an average energy gain of $\mathcal{E} = 4$ per visit would increase some particle energies by $4^5 = 1024$. This is the main idea underlying the sequential particle-acceleration model described next. 

\section{Sequential Acceleration in Multiple Accelerators} 
\label{sec:PALMI}

\subsection{Initial Distribution Function}
\label{sec:maxwellian_A}

We assume that the ambient corona is characterized by a Maxwellian particle distribution function at temperature $T$ and with an isotropic distribution in pitch angle, $f_0(E,\theta)= f_0(E)/90^\circ$. The fractional number of particles with energies in the range $(E,E +dE)$ is
\begin{eqnarray}
  \label{eq:dens_dist}
   f_0(E) dE & = &  \frac{2}{\sqrt{\pi}}   e^{- \left(\frac{E}{  k_{B} T}\right)}  \sqrt{ \left( \frac{E}{k_{B} T} \right)} \frac{dE}{k_{B} T} ,
 \end{eqnarray}
where $k_{B}$ is the Boltzmann constant. 

In terms of the dimensionless kinetic energy, defined as $\overline{E} = \frac{E}{k_{B} T}$, the initial distribution is 
\begin{eqnarray}
  \label{eq:unitless}
   f_0(\overline{E}) & = & \frac{2}{\sqrt{\pi}} e^{- \overline{E} }  \sqrt{\overline{E}}. 
\end{eqnarray}

\begin{figure}[ht]
   \includegraphics[width=\columnwidth]{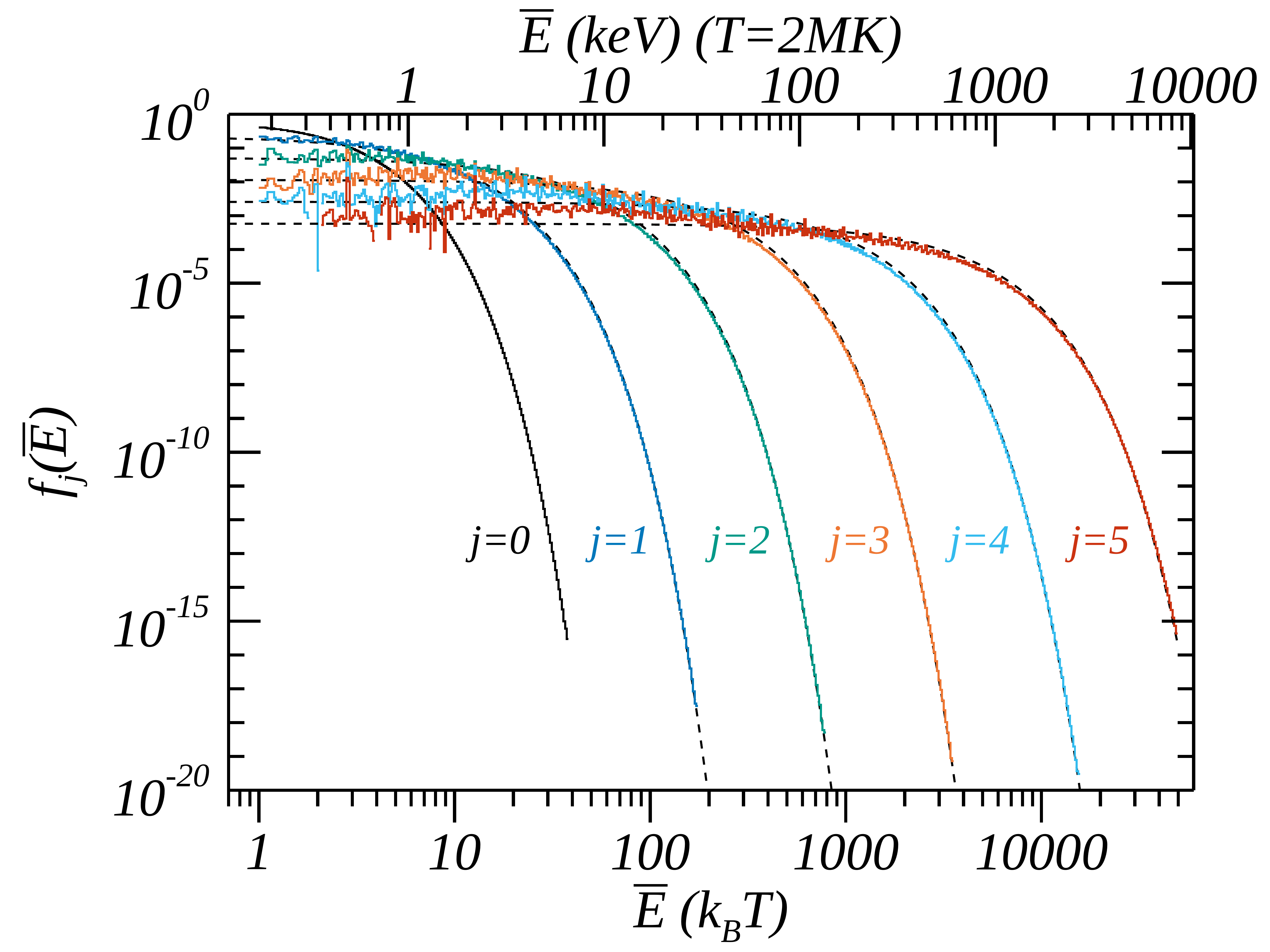}
   \caption{Normalized energy spectra for different cycles (cycle number $j$ is shown next to each colored curve) of A2 accelerators. The units of the bottom horizontal axis (logarithmic bins of size $\log(1.02)$) are $k_{B}T$, where $T$ is the ambient coronal temperature of $f_0$, and the top energy axis is in keV, for an assumed temperature $T = 2$ MK. (This double-unit horizontal-axis setup continues in subsequent figures.) Black dashed lines show the corresponding fitted functions $e^{-\overline{E}/R}/R$.}
   \label{fig:each_cycle}
\end{figure}

Spectrum $f_0(\overline{E})$ is shown as the solid black curve in Figure \ref{fig:each_cycle} (labeled $j=0$). For easier comparison to observations, the top horizontal axis of the figure shows energy in keV for an assumed background temperature $T = 2$ MK. 

To numerically track the particle energies and pitch angles as they evolve in time inside an accelerator, we represent the $\overline{E}$-$\theta$ phase space with a 2D grid of energy and angle bins. The range of $\overline{E}$ is from $0$ to $\overline{E}_{m} = 50,000$ (appropriately large to study high-energy acceleration), and $0 \leq \theta \leq 90^\circ$. Each dimension is binned at regular intervals, $\Delta {\overline{E}}=0.1$ and $\Delta \theta = 0.1^\circ$. 

We will refer to the fractional number of particles in the energy range $(\overline{E}_{j},\overline{E}_{j}+\Delta {\overline{E}})$ and pitch angle range $(\theta_{k}, \theta_{k}+\Delta \theta)$ as a ``macroparticle'' $N(j,k)$. The initial macroparticle distribution is  
\begin{eqnarray}
  \label{eq:number_E}
    N_0(j,k) & =  &  \frac{\Delta \theta}{90^\circ} \int_{\overline{E}_{j}}^{\overline{E}_{j}+\Delta {\overline{E}}}  f_0(\overline{E}) d\overline{E}   \\ \nonumber
    & =  & \frac{\Delta \theta}{90^\circ} \left[\mathcal{N}(\overline{E}_{j}+\Delta \overline{E}) - \mathcal{N}(\overline{E}) \right], 
\end{eqnarray}
where $\mathcal{N}(\overline{E})$ is the normalized number of particles between energies $0$ and $\overline{E}$ given by 

\begin{eqnarray}
  \label{eq:tot_numb}
   \mathcal{N}(\overline{E}) & =  & \erf{\left(\sqrt{\overline{E}}\right)} - \frac{2}{\sqrt{\pi}} \sqrt{\overline{E}} e^{-\overline{E}}
\end{eqnarray}
and $\erf$ is the error function.

\begin{figure}
   \centering
   \begin{center}$
     \begin{array}{cc}
        \includegraphics[trim={2cm 9cm 1cm 8.5cm},clip,width=\columnwidth]{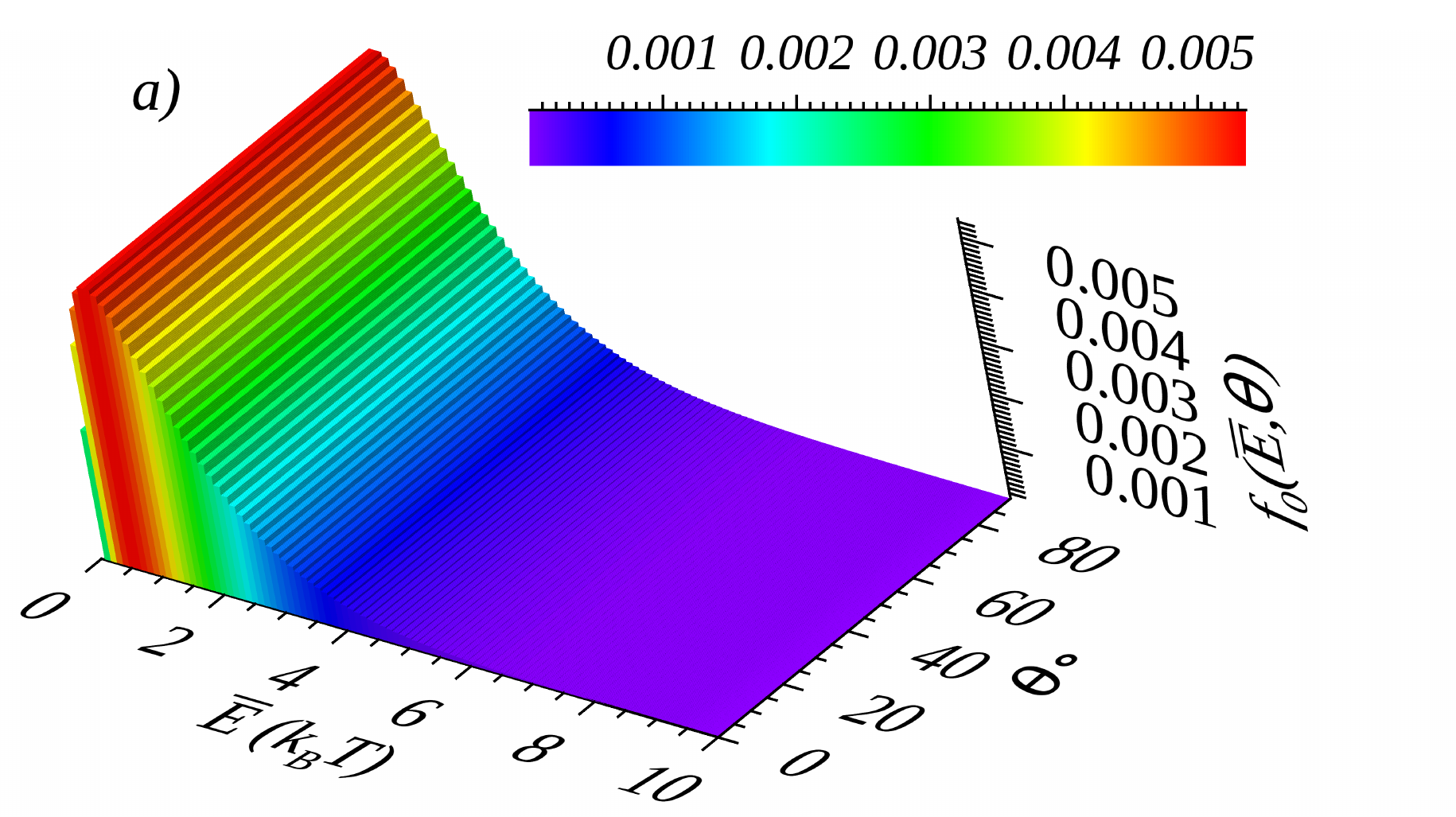}\\
        \includegraphics[trim={2cm 9cm 1cm 9cm},clip,width=\columnwidth]{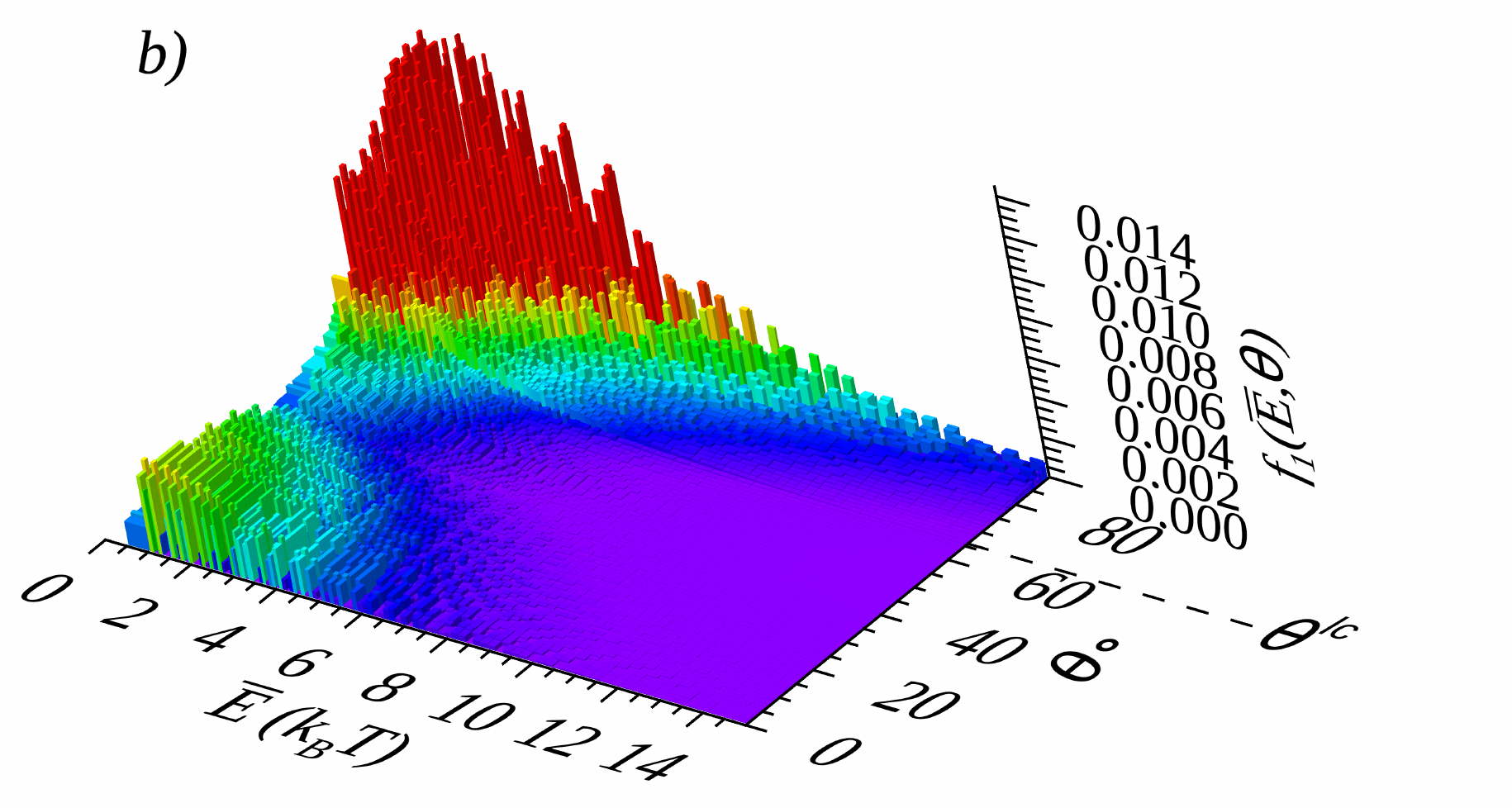}\\
        \includegraphics[trim={1.9cm 9.2cm 1cm 9cm},clip,width=\columnwidth]{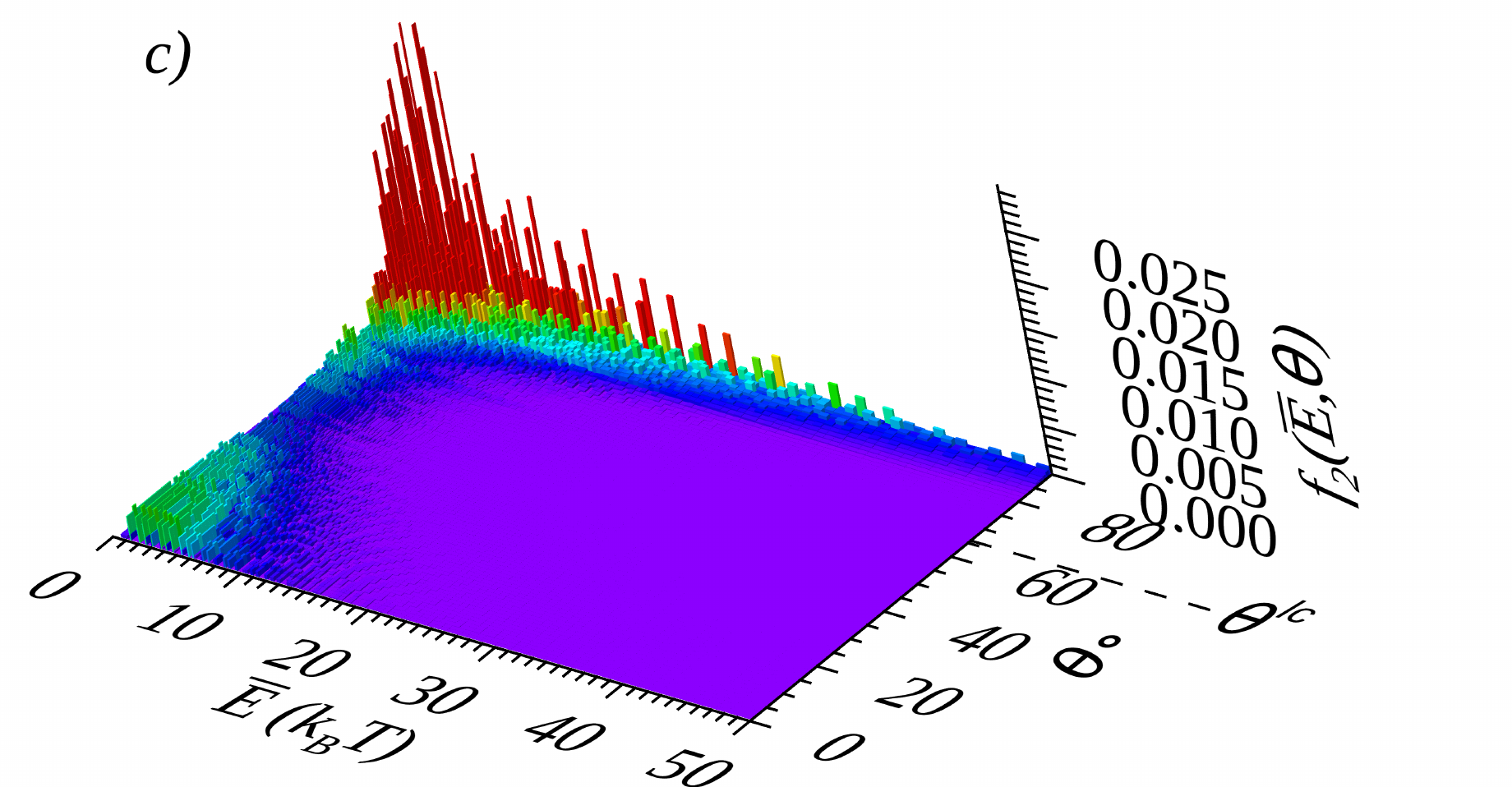}
     \end{array}$
   \end{center}
   \caption{Normalized particle distributions $f_j(\overline{E}, \theta)$ for A2 accelerators, saturated at the maximum of the color table. a) $f_0$, b) $f_1$, c) $f_2$. In panels b) and c), data have been resampled to logarithmic bins of size $\log(1.02)$ and $\Delta \theta=1^\circ$. $\theta_{\hbox{lc}}$ (which is the same for accelerators of the same type) is shown with a horizontal dashed line. 
   }
   \label{fig:Init_energy_distrib_ff}
\end{figure}

The initial isotropic distribution of macroparticles $f_0(E,\theta)$is shown in Figure \ref{fig:Init_energy_distrib_ff}a. Those macroparticles with $\theta > \theta_{\hbox{lc}}$ 
are mirroring; the rest are transiting. 

\subsection{Macroparticle Evolution in One Accelerator}
\label{sec:oneaccel}

At the end of the lifetime of an accelerator, each macroparticle initially in $f_0(\overline{E},\theta)$ at $\overline{E}_i, \theta_i$ will have a final $\overline{E}_f, \theta_f$ determined by the method described in the previous section (Fig.\ \ref{fig:energization_rules}.) The macroparticle is assigned to the location on the 2D grid closest to $\overline{E}_f, \theta_f$. Hence, at this final time, each grid cell in the $\overline{E}$-$\theta$ phase space may have one, several, or no macroparticles. Those particles that achieve energies larger than $\overline{E}_{m}$ are lost, but this is a negligible number in our calculations. In this section, several figures present results for A2, which has the largest energy gains and the largest proportion of mirroring population. Similar conclusions were drawn for A1 but are not shown.

We estimated the distribution of macroparticles at the end of the lifetime of the accelerator $f_1(\overline{E}, \theta)$ by summing macroparticles inside each cell of the $\overline{E}$-$\theta$ phase space. $f_1(\overline{E}, \theta)$, shown in Figure\ \ref{fig:Init_energy_distrib_ff}b, has the same total number of particles $N$ (normalization factor for all distributions) as $f_0$, redistributed across the grid. The spikiness of $f_1$ is due to the discretization of the phase space: some of the bins do not have macroparticles at this particular time. 

$f_1$ is highly anisotropic in pitch angle. Particles with large $\theta$ have larger final energies than their counterparts at small $\theta$. All of the macroparticles have increased their pitch angle $\theta$, consistent with the sharp initial slope of $\theta_f$ (dashed red curve) in Figure \ref{fig:energization_rules}. Both of these features reflect the dominant role of betatron acceleration, which strongly increases the energy of motion perpendicular to the magnetic-sfield direction as the field strength increases. 

Some macroparticles switch from mirroring to transiting populations: the percentage of mirroring particles in $f_1$ is $58\%$, as opposed to $74\%$ in $f_0$. This is due to the reduction in the magnetic mirror ratio of the accelerator as it evolves from highly elongated to nearly circular. Nevertheless, for A2, mirroring particles in $f_1$ are the largest population and possess the highest energies. In contrast, the opposite is true for A1: its mirroring percentages are $21\%$ ($f_1$) and $38\%$ $(f_0)$, and the energies are highest for transiting particles. This reflects the less prominent role of betatron acceleration for A1, whose magnetic-field compression is much less than that of A2.

A2's $f_1$ has more particles at high energies than $f_0$. Its energy spectrum is shown as the blue curve ($j=1$) in Figure \ref{fig:each_cycle}. The average particle energy $\left<\mathpunct{\overline{E}} \right>$ has increased from $1.5$ in $f_0$ to $4.3$ in $f_1$, nearly a factor of three. This energy increase is modest but not insignificant. In the next section, we examine the consequences of having particles ``visit'' several accelerators sequentially, receiving a boost in energy at each stage. 

\subsection{Sequential Accelerators}
\label{sec:visiting}

To investigate the effect of sequential accelerators of the same type on the particle distribution, we take the final distribution $f_1$ from the single-accelerator experiment above and evolve it using the same rules used to evolve $f_0$ into $f_1$. Processing $f_1$ through the same type of accelerator results in a new final distribution function $f_2$, which has more particles at higher energies and a more anisotropic pitch-angle distribution than $f_1$. For example, $f_2$ for A2 accelerators is shown in Figure \ref{fig:Init_energy_distrib_ff}c. 

This process is repeated sequentially multiple times, yielding a distribution function $f_j$ after $j$ cycles. During each cycle, the total number of particles in each $f_j$, $N$, is conserved, but the number of particles at high (low) energies increases (decreases) as the particles are accelerated, and ever more particles achieve large pitch angles, resulting in an increasingly anisotropic distribution.  

\begin{figure}[ht]
   \includegraphics[width=\columnwidth]{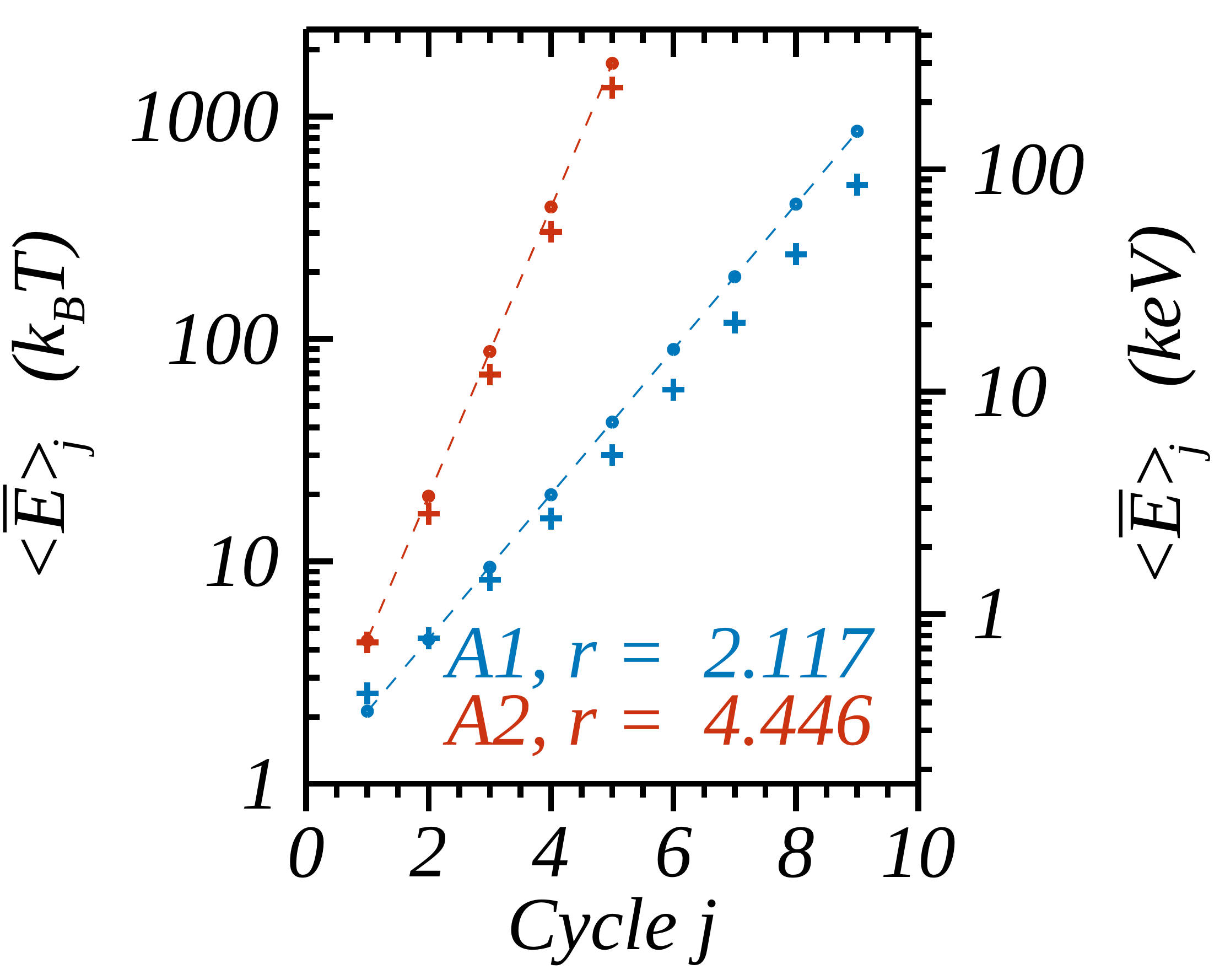}
   \caption{Average energy of cycled distributions. Crosses show $\left <\overline{E}\right>$ of $f_j$ as functions of cycle $j$ for accelerators A1 (blue) and A2 (red). $\left<\overline{E}\right>_0 = 1.5$ for both accelerators (not shown). The left vertical axis is in units of $k_{B}T$, where $T$ is the ambient coronal temperature of $f_0$; the right energy axis is in keV, for an assumed temperature $T = 2$ MK. (This double-unit vertical-axis setup continues in subsequent figures.) Circles show $R$ from fitting $f_j$'s spectra with function $e^{-\overline{E}/R}/R$ (e.g., see Figure\ \ref{fig:each_cycle}), for which $R = \left<\overline{E}\right>$ (See \S \ref{sec:analytical}). Dashed lines show $R$ fitted as $r^j$ ($j\geq1)$. The fitted $r$ is shown in the color-coded annotations.
   } 
   \label{fig:full_cycle_energy_gain}
\end{figure}

The energy spectra for the particles that gain energy by sequentially visiting A2 accelerators are shown in Figure \ref{fig:each_cycle} with colored lines, up to $j=5$ cycles. In later cycles, there are large fluctuations in the distributions at low energies because not many particles are left in that energy range. Every new cycle has a spectrum with more high-energy particles and higher average energy than the previous one. The last distribution in the sequence, $f_5$, presents a very hard spectrum with a small spectral index. We find that the exponential functional form $e^{-\overline{E}/R}/R$ fits all of the distributions reasonably well, as shown in Fig. 4 (dashed black lines); we will make use of this form in our analytical treatment in \S\ref{sec:analytical}. 

$\left <\overline{E}\right>_j$ of each cycle increases with the number of cycles, as shown with crosses in Figure \ref{fig:full_cycle_energy_gain} for sequences of accelerators A1 (blue) and A2 (red). Circles show $R$ from fitting $f_j$'s spectra with function $e^{-\overline{E}/R}/R$, for which $R = \left<\overline{E}\right>$.

It is unrealistic to expect that all of the particles in any accelerator will be transferred to and cycled through a new accelerator. A more plausible scenario is that some fraction of each accelerator's population will be transferred to a new accelerator, to participate in another round of energization in a succeeding cycle. This is the basis for the model discussed in the next section. 

\begin{figure}[ht]
   \includegraphics[width=\columnwidth]{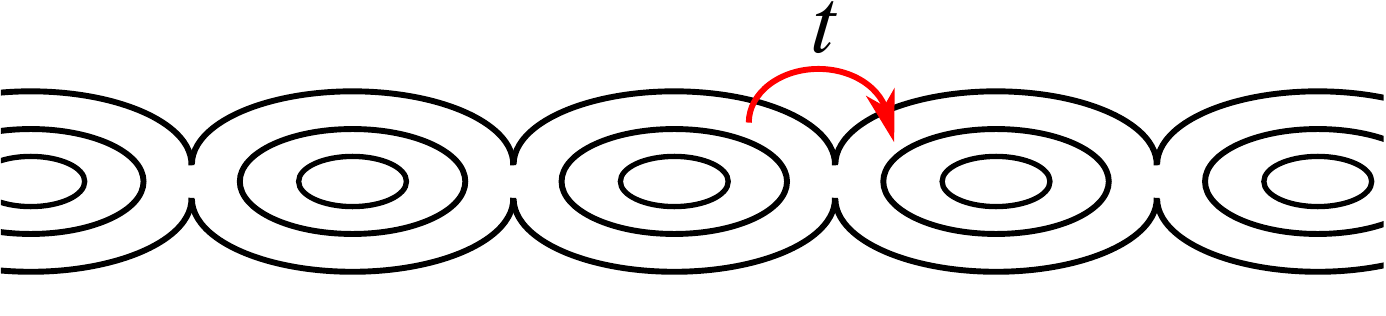}
   \caption{Cartoon depicting a sequence of islands and particles being transferred between them. $t$ is the fraction of particles transferred from one accelerator to another one (parameter of the model).}
   \label{fig:island_sequence_cartoon}
\end{figure}

\subsection{Transfer of Particles between Accelerators}
\label{sec:transfer_factor}

We generalize the cycling algorithm presented above by transferring only a fraction of the particles from the preceding to the following accelerator and by allowing the new accelerator to entrain ambient particles along with the previously accelerated particles. This emulates a continuously reconnecting flare current sheet in which new islands form that contain fresh coronal plasma but that also capture energetic particles that have escaped a previously formed island.

To represent the fraction of particles from one accelerator transferred to the next accelerator (see cartoon in Fig.\ \ref{fig:island_sequence_cartoon}), we define a typical transfer factor, $t \leq 1$. We assume that $t$ is the same for all accelerators. For simplicity, we further assume that particles at all energies are equally likely to be transferred from island to island and that the particles' pitch angles in the new accelerator are the same as in the preceding one. As we show below, all of these simplifying assumptions allow us to make analytical progress in calculating the evolving particle distribution function. 

As before, if the first accelerator in the sequence has an initial distribution $f_0$ (Figure \ref{fig:Init_energy_distrib_ff}a), after one cycle its final distribution is $f_1$ (e.g., A2's $f_1$ is shown in Figure \ref{fig:Init_energy_distrib_ff}c). We express this result in the form 
\begin{align}
   \label{eq:first_cycle}
   f_i^{(1)} &= f_0, \\
   f_f^{(1)} &= f_1,
\end{align}
where the subscripts $i,f$ represent the initial and final distributions and the superscript $(1)$ indicates the first cycle in the model sequence. The subsequent accelerator will have an initial distribution that is characterized in part by the background distribution $f_0$, plus a fraction $t$ of the previously accelerated distribution $f_1$. 

We express the initial distribution function for the second accelerator in the form \begin{align}
   \label{eq:second_cycle_i}
   f_i^{(2)} &= (1-t) f_0 + t f_f^{(1)} \\
             &= (1-t) f_0 + t f_1
\end{align}
$f_i^{(2)}$ has lost a fraction $t$ of background particles and gained a fraction $t$ of $f_1$. For $t<<1$, $f_i^{(2)}$ deviates slightly from an isotropic Maxwellian distribution. If this population is now cycled through accelerators of the same type following the prescribed rules from \S \ref{sec:previous} (Fig.\ \ref{fig:energization_rules}), each component distribution, $f_0$ and $f_1$, will evolve to the next cycled distribution, $f_1$ and $f_2$ (e.g., A2's $f_2$ is shown in Figure \ref{fig:Init_energy_distrib_ff}c), respectively. The final distribution then will be 
\begin{equation}
   \label{eq:second_cycle_f}
    f_f^{(2)} = (1-t) f_1 + t f_2. 
\end{equation}
For $t<<1$, $f_f^{(2)}$ deviates slightly from the anisotropic distribution $f_1$.  

This process can be applied recursively, prescribing that at each cycle a fraction $t$ of the preceding accelerator's population is transferred to the new accelerator. We tacitly assume that the particles are collisionless, so populations do not interact over the time scale of the full acceleration process. Consequently, each component distribution $f_j$ evolves separately to $f_{j+1}$. 

We summarize this procedure in Table \ref{tab:table1}. The final distribution after $n$ cycles is the linear combination of component distributions 
\begin{eqnarray}
   \label{eq:overlinedist}
    f_f^{(n)}(n,t) &=& \left(\dfrac{1-t}{t} \right) \sum \limits_{j=1}^{n} t^{j} f_{j} + t^{n}f_{n} 
    \label{eq:overlinedist1}
\end{eqnarray}
The last contribution is negligible at low energies when $n$ is large. $f_f^{(n)}$ has the same total number of particles $N$ as each of the cycled distributions $f_j$.

We constructed $f_f^{(n)}$ for sequences of accelerators of the same type (A1 or A2) for different transfer parameters $t$ and cycle numbers $n$. To reduce computer memory usage, each $f_j(\overline{E}, \theta)$ in Equation \ref{eq:overlinedist1}
was resampled into logarithmic bins of size $\log(1.02)$ and $\Delta \theta=1^\circ$ (e.g., Figures \ref{fig:Init_energy_distrib_ff}b,c). The chosen values of $t$ range from $10^{-5}$ to $0.7$, arranged in multiples of 10 of the triplet $(1,5,7)\times10^{-5}$. We study the resulting spectra of the sequential final distribution functions in the next section. 

\begin{figure}[ht]
   \includegraphics[width=\columnwidth]{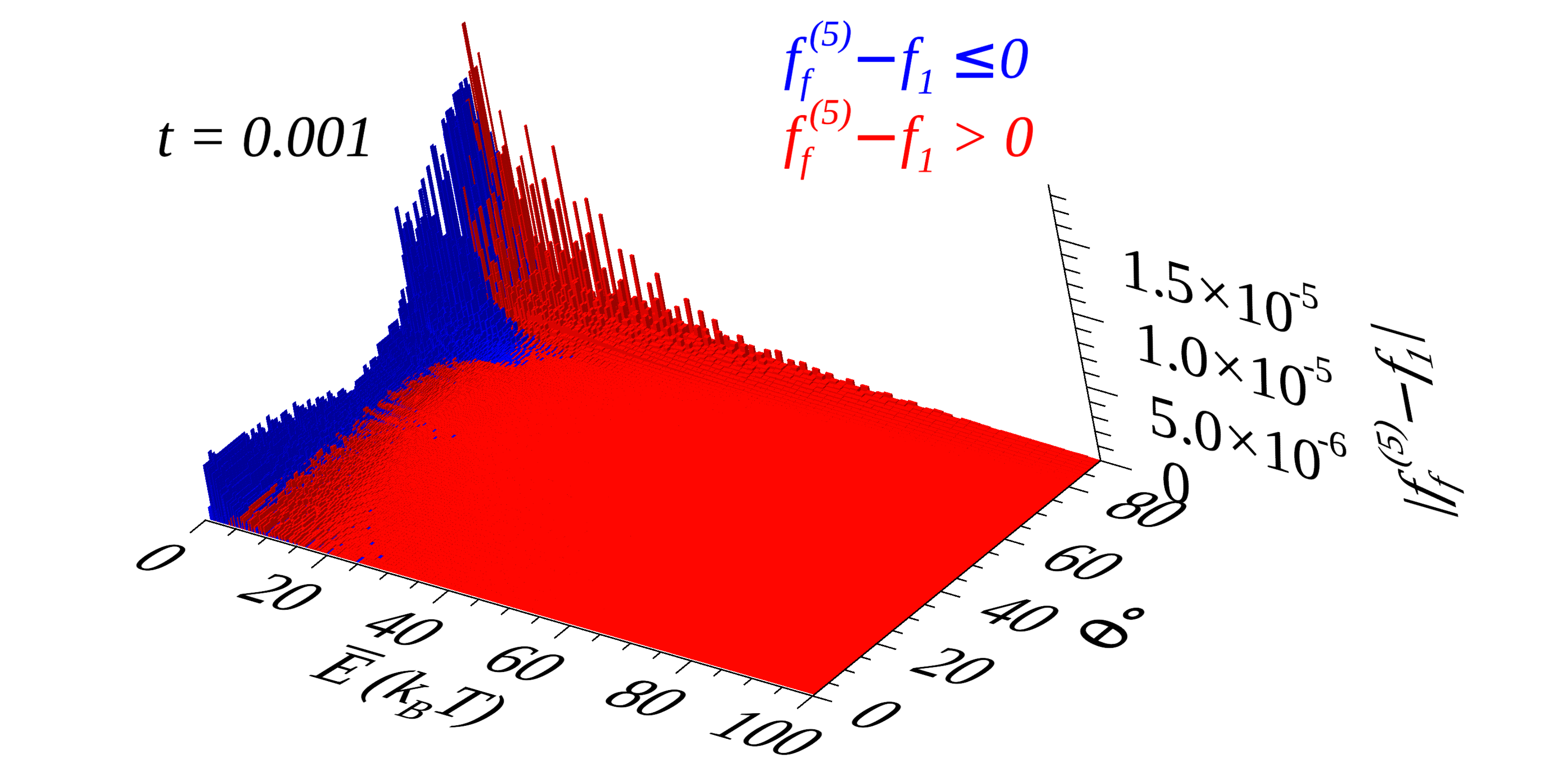}
   \caption{Absolute value of the difference between the sequential final distribution $f_f^{(5)}$ (Equation \ref{eq:overlinedist1}) and $f_1$ (Figure \ref{fig:Init_energy_distrib_ff}b) for accelerator A2 with $t=0.001$. Red (blue) color indicates positive (negative) difference.}
   \label{fig:difference_fn_f1}
\end{figure}

For small $t$, $f_f^{(n)}$ resembles $f_1$, except for a small increase of particles at high energy and large pitch angle at the expense of a loss of particles at low energies (see example of the differences between these distributions in Figure \ref{fig:difference_fn_f1} for $n=5$, $t=0.001$ and accelerator A2.) 

\begin{table*}[ht]
  \begin{center}
    \caption{Model distribution functions.}
    \label{tab:table1}
    \begin{tabular}{|c|l|l|}
    \hline
      \textbf{Cycle} & \textbf{Initial Distribution} & \textbf{Final Distribution}\\
      \hline
       \hline
       \tablespaceaddtwo
      $1$ & $f_i^{(1)} = f_0$ & $f_f^{(1)} = f_1$ \\ [1ex]
      \hline
      \tablespaceaddtwo
      $2$ & $f_i^{(2)} = (1-t)f_0 + t f_f^{(1)} = (1-t)f_0 + t f_1 $ & $ f_f^{(2)} = (1-t)f_1 + t f_2 = f_1 + t (f_2 -f_1) $ \\  [1ex]
      \hline
      \tablespaceaddtwo
      $3$ & $f_i^{(3)} = (1-t)f_0 + t f_f^{(2)} = (1-t)f_0 + t \left[(1-t) f_1 +  t f_2 \right] $ & $ f_f^{(3)} = (1-t) f_1 + (1-t) t f_2 +  t^2f_3  $  \\
       &  & $ \qquad  = f_1 + t (f_2 -f_1) + t^{2} \left(f_3 - f_2 \right) $  \\  [1ex]
      \hline
      \tablespaceaddtwo
      $\vdots$ & $\vdots$  & $\vdots$ \\ [2ex] 
      \hline
     \tablespaceaddtwo
            $n$ & $f_i^{(n)} = (1-t)f_0 + t f_f^{(n-1)} $ & $f_f^{(n)} = f_f^{(n-1)} + t^{n-1}\left(f_{n}-f_{n-1}\right) $  \\ 
             & \textcolor{white}{$f_i^{(n)}$} $= (1-t) \sum \limits_{j=0}^{n-1} t^{j} f_{j} + t^{n}f_{n-1} $ & \textcolor{white}{$f_f^{(n)}$}
            $=\left(\dfrac{1-t}{t} \right) \sum \limits_{j=1}^{n} t^{j} f_{j} + t^{n}f_{n} $  \\
       [1ex] 
      \hline
    \end{tabular}
  \end{center}
\end{table*}

\section{Spectra}
\label{sec:results}

\begin{figure}[ht]
   \includegraphics[width=\columnwidth]{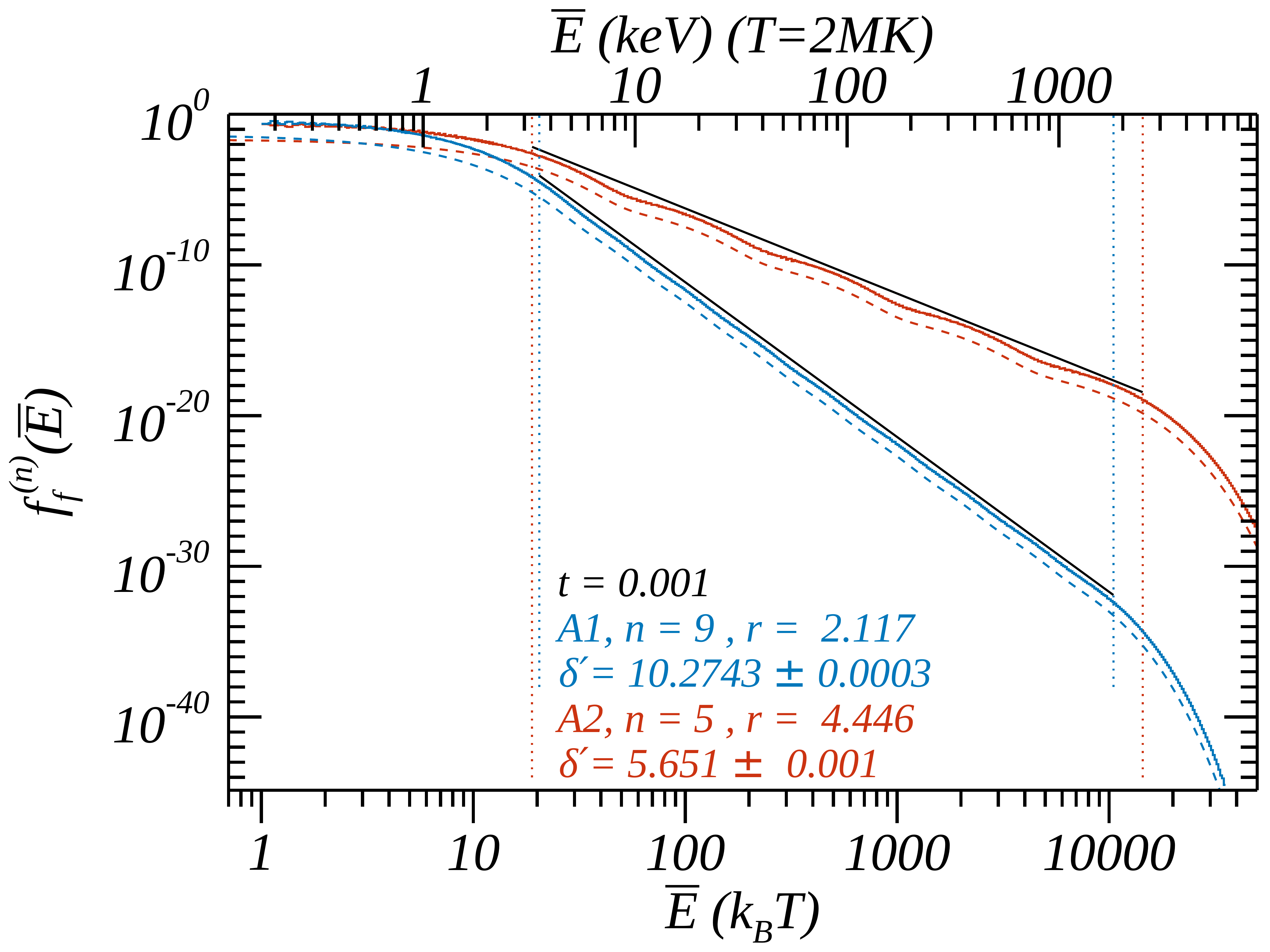}
   \caption{Energy spectra for sequences of accelerators A1 (blue) and A2 (red) with transfer factor $t=0.001$ (solid curves = simulated data; dashed curves = analytical function, Equation \ref{eq:dist_anal_form} in \S \ref{sec:analytical} with $\alpha = 0$, and annotated efficiency $r$ and final cycle $n$), shifted downward by a factor of 10 for visual clarity. Black lines have slopes equal to the fitted color-coded spectral indices $\delta'$, whose estimated uncertainties are given. Fitted $E_{\hbox{leb}}$ (left) and $E_{\hbox{heb}}$ are shown with color-coded vertical dotted lines.}
   \label{fig:mirror_transit_index}
\end{figure}

We calculated the energy spectra of all our simulated $f_f^{(n)}$ by summing over pitch angle. In general, the spectrum after a few cycles has the following features: 1) a flat spectrum at low energies, consisting mainly of contributions $f_j$ with small $j$, 2) a power-law-like shape $\sim E^{-\delta'}$ at intermediate energies, and 3) a rapid decrease at high energies. Examples are shown in Figure \ref{fig:mirror_transit_index} for  accelerators A1 (blue) and A2 (red), both for the transfer parameter $t=0.001$. The number of cycles used is $n=9$ for A1 and $n=5$ for A2: these numbers were found to yield similar power-law energy ranges for the two accelerators. Because A1 is less efficient at accelerating particles than A2, more cycles are required to produce similar high-energy breaks. As expected, A2 presents the hardest power law.

A smooth transition between the Maxwellian-like distribution at low energies and the power-law region of the spectrum shown in Figure \ref{fig:mirror_transit_index} supplants the usually assumed low-energy cutoff where the power-law distribution ends abruptly. In the next section, we will estimate the energy above which the distribution can be well approximated as a power law, which we denote the \textit{low-energy break} $E_{\hbox{leb}}$. We note that the transition is smooth and, hence, there is no well-defined precise value for this energy. 

The middle, power-law-like region of the spectrum is gently modulated due to small-amplitude bumps associated with the discrete cycles(Figure \ref{fig:mirror_transit_index}). Although A2's distribution is more sinuous than A1's, both curves are fit well by power laws, using the method explained in the Appendix. 

\begin{figure}[ht]
   \includegraphics[width=\columnwidth]{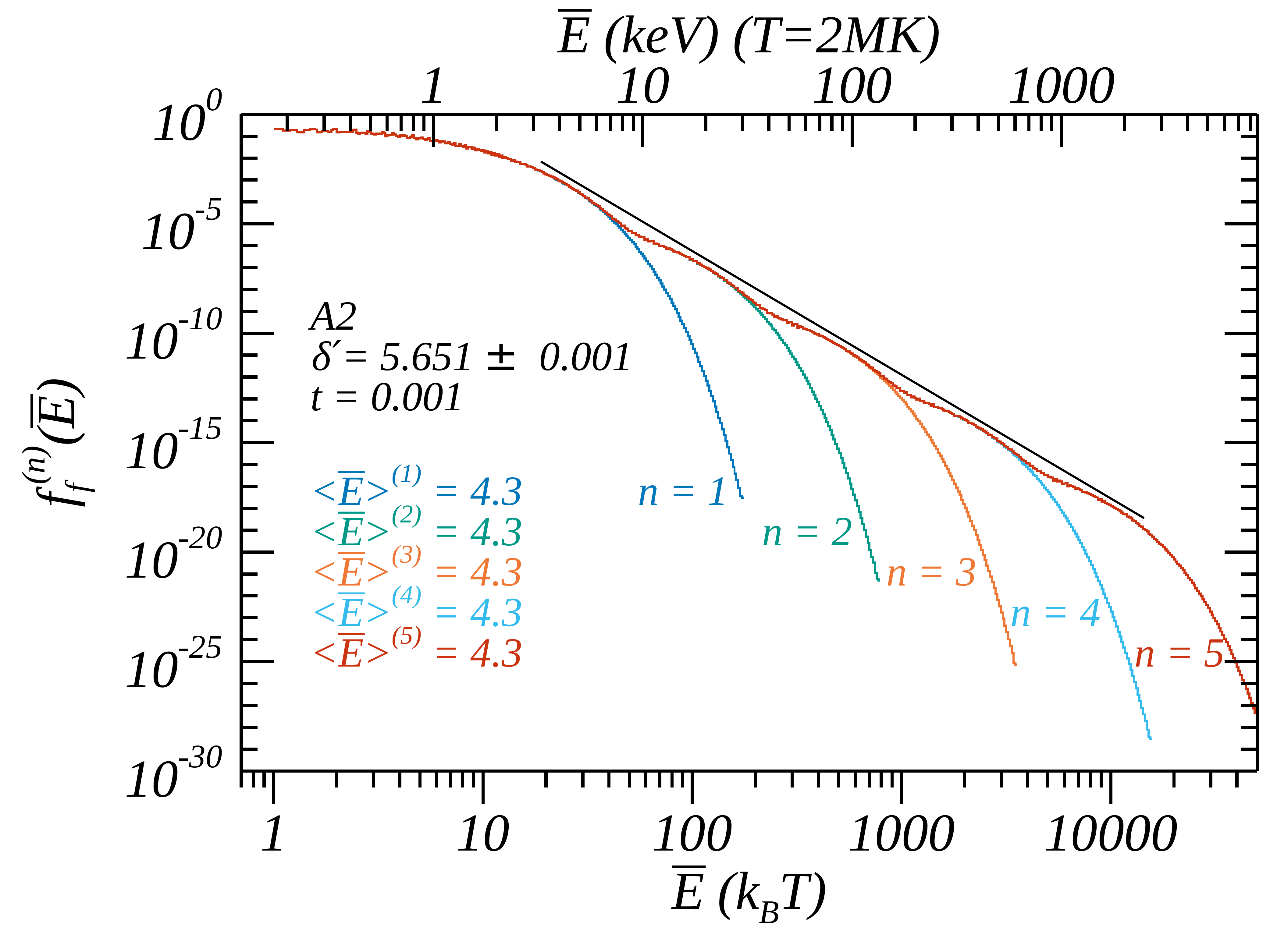}
   \caption{Each color-coded line shows $f_f^{(n)}$ for different $n$ (annotated) and fixed $t=0.001$ for a sequence of A2 accelerators. The final distribution $f_f^{(5)}$ (red) overlaps the other distributions except at high energies. The average energy of each distribution is shown with color-coded annotations. The black straight line is the fitted power law $\sim E^{-\delta'}$  ($\delta'$ is annotated in black) for the case $n=5$, plotted between fitted $E_{\hbox{leb}}$ and $E_{\hbox{heb}}$ for $f_f^{(5)}$. This line is shifted upward by $10^{0.5}$ for visual clarity.}
   \label{fig:high_energy_break}
\end{figure}

Each sequential cycle extends the range of energies for which the spectrum shows a power-law shape, i.e., the \textit{high-energy break} $E_{\hbox{heb}}$ increases with the number of visited accelerators. The tail after $E_{\hbox{heb}}$ has approximately the shape of an exponential decay and corresponds to the last terms of the sequence in Equation \ref{eq:overlinedist} and Table \ref{tab:table1}.  Examples of A2's spectra are shown in Figure \ref{fig:high_energy_break} for different final cycles $n$ (color-coded) with transfer factor $t=0.001$. Only $n = 5$ accelerators and a particle transfer factor $t=0.001$ were needed to increase the energies of some particles by two orders of magnitude and form a power law.

Three features of the distribution do not change much as the number of cycles increases. First, $E_{\hbox{leb}}$ is essentially set by the initial cycle and changes little for additional cycles. Second, as shown in Figure \ref{fig:high_energy_break}, the spectral index does not change significantly as the number of visited accelerators increases. Third, as indicated in the annotations, $\left <\overline{E}\right>$ is nearly invariant. The process does not add much energy to the system, because only a very small fraction of energized particles is transferred to the next accelerator. The average energy is essentially that of $f_1$, i.e., it is dominated by the acceleration of the ambient Maxwellian particles in $f_0$. 

We emphasize that the transfer of particles between accelerators is assumed to be uniform across all energies. Therefore, the large number of high-energy particles is not an artifact of particle-acceleration or transfer mechanisms that favor particles with high energies. The number of particles at lower/higher energies decreases/increases with each cycle, redistributing the population from one cycle to the next in such a way that the area under the curve and the average energy remain nearly unchanged throughout. 

To determine $E_{\hbox{leb}}$ and $E_{\hbox{heb}}$, as well as spectral indices of final distributions, we modeled the central region as a power law $C E^{-\delta'}$, where $C$ is a normalization constant. To estimate these parameters and their uncertainties, we developed an automatic curve-fitting procedure that requires minimal human intervention, as described in the Appendix. Examples of fitted power laws for $t = 0.001$ are shown in Figures \ref{fig:mirror_transit_index} and \ref{fig:high_energy_break} (black solid lines). 

\begin{figure}[ht]
   \includegraphics[width=\columnwidth]{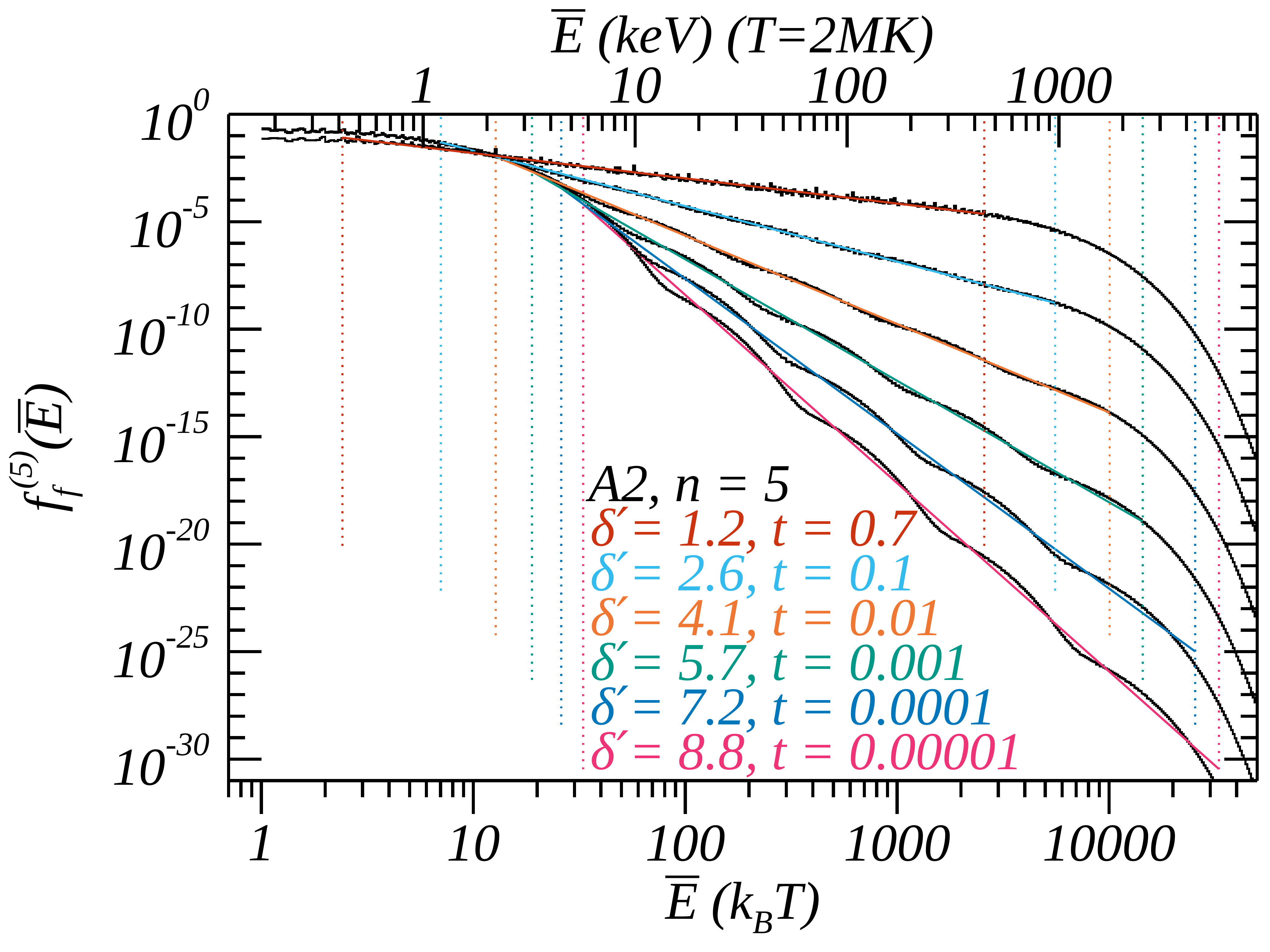}
   \caption{Spectral index as a function of $t$. A2 distributions with $n = 5$ are shown in black and their fitted power-law curves (color-coded solid lines) for several values of the transfer factor $t$ (color-coded annotations). Color-coded vertical dotted lines show fitted low-energy breaks $E_{\hbox{leb}}$ (left) and high-energy breaks $E_{\hbox{heb}}$ (right).}
   \label{fig:transfer_factor_effect}
\end{figure}

\subsection{Dependence of Fitted Parameters on Transfer Factor $t$}
\label{sec:fitsection}

Previously, we presented results for accelerators A1 and A2 using the fixed value $t = 0.001$ for the transfer factor $t$. 
The larger the transfer factor, the greater the number of particles that are transferred from one accelerator to the next. Hence, we expect more energetic particles and harder spectral indices in the distributions with larger $t$. Figure \ref{fig:transfer_factor_effect} illustrates these effects for  A2 with $n=5$. In addition, as $t$ increases, we find that the bumps in the distribution become less pronounced. As explained in \S \ref{sec:analytical}, this occurs because the weight of each cycle on the overall curve decreases. 

\begin{figure}[ht]
   \includegraphics[width=\columnwidth]{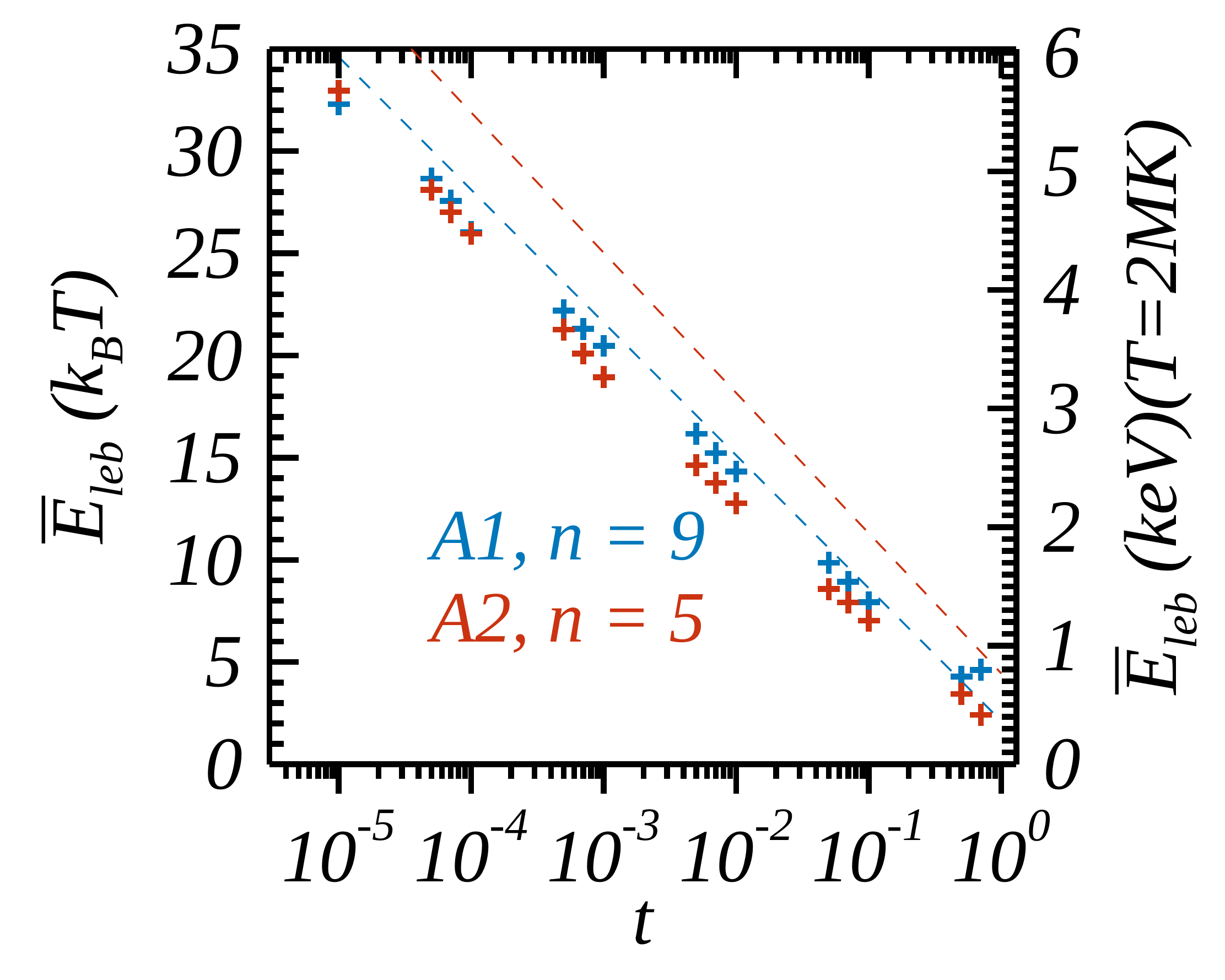}
   \caption{$E_{\hbox{leb}}$ as a function of $t$. Crosses show fitted low-energy break $E_{\hbox{leb}}$ for A1 (blue) and A2 (red) as functions of the transfer factor $t$ for fixed $n$ (annotated). The average percent error is $<2$\%. Color-coded dashed lines are theoretically predicted values of $E_{\hbox{leb}}$ (Equation \ref{eq:dist_anal_max_6} in \S \ref{sec:analytical}, with the color-coded annotated $r$ shown in Figure\ \ref{fig:full_cycle_energy_gain} and with  $\alpha=0$).}
   \label{fig:lowenergybreak}
\end{figure}

A visual inspection of Figure \ref{fig:transfer_factor_effect} suggests that all of the distribution functions converge to a single point near $\overline{E} = 15$, which might imply a common low-energy break for all the curves. However, fitted $E_{\hbox{leb}}$ and $E_{\hbox{heb}}$ values (color-coded vertical dotted lines in the figure) decrease with transfer parameter $t$. Fitted low-energy breaks for A1 (blue) and A2 (red) are shown with crosses as functions of the transfer factor $t$ and fixed $n$ in Figure \ref{fig:lowenergybreak}. 

Interestingly, although the low-energy breaks change with transfer factor, they are quite similar for the two accelerators. The reason for this weak dependence is explained in \S \ref{sec:analytical}. The curves have an approximate logarithmic dependence on $t$, with the low-energy breaks decreasing as the number of particles transferred between accelerators increases and the number of particles in the power-law range increases. The range in low-energy breaks is small, varying over about 1 to 6 keV for an assumed background temperature of $2$ MK (larger background temperatures would increase the low-energy breaks.)

\begin{figure}[ht]
   \includegraphics[width=\columnwidth]{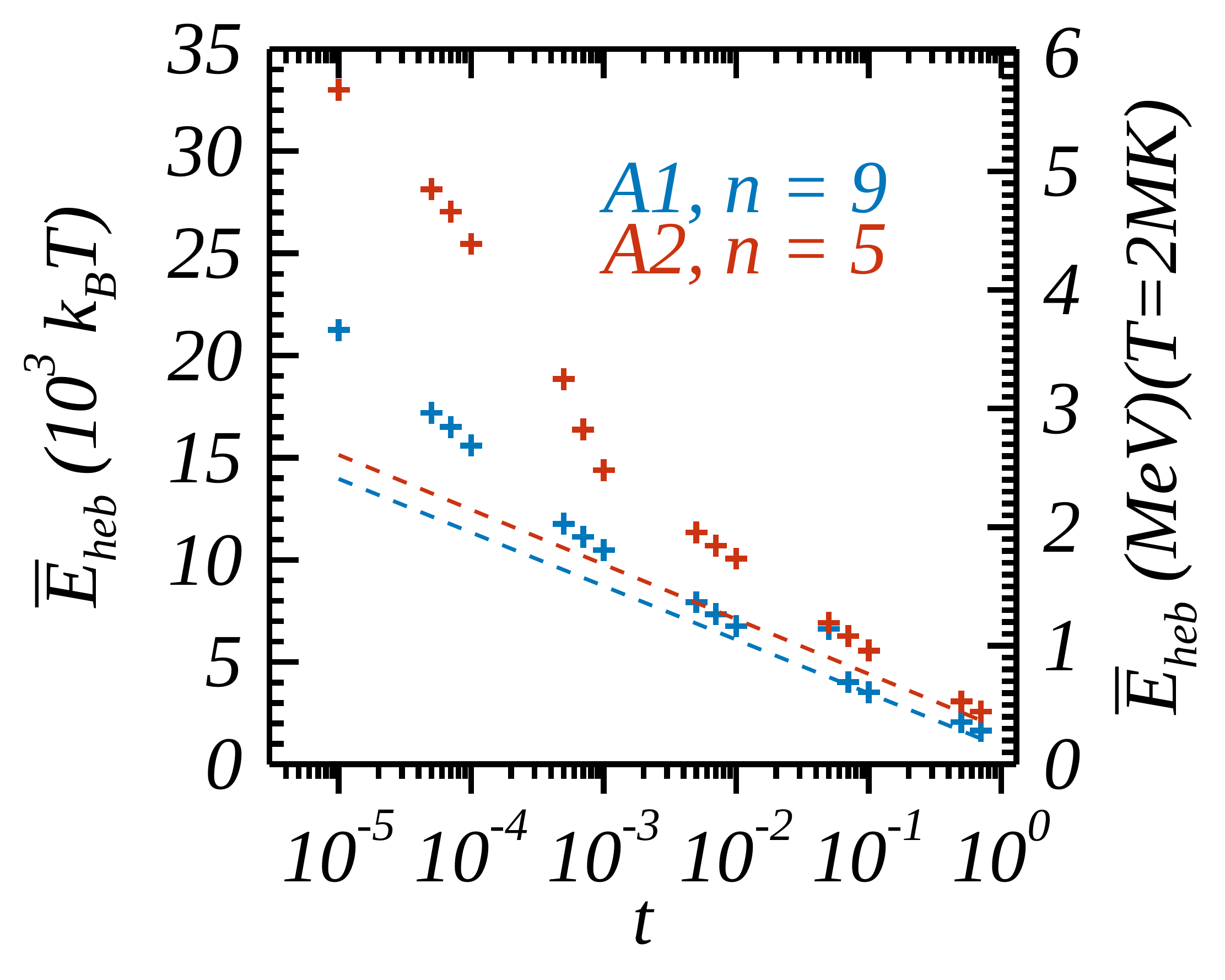}
   \caption{$E_{\hbox{heb}}$ as a function of $t$. Crosses show fitted high-energy break $E_{\hbox{heb}}$ for A1 (blue) and A2 (red) as functions of the transfer factor $t$ for fixed $n$ (annotated). The average percent error is $<2.3$\%. Color-coded dashed lines are theoretically predicted values of $E_{\hbox{heb}}$ (Equation \ref{eq:dist_anal_max_9} in \S \ref{sec:analytical}, with the color-coded annotated $r$ shown in Figure\ \ref{fig:full_cycle_energy_gain} and with  $\alpha=0$).}
   \label{fig:highenergybreak}
\end{figure}

We found a similar decreasing trend for the fitted high-energy breaks as functions of the transfer factor $t$, plotted in Figure \ref{fig:highenergybreak} for A1 (blue) and A2 (red). These curves show more pronounced differences between the accelerators than those in Figure \ref{fig:lowenergybreak}. The high-energy break shifts to lower energies as the transfer factor increases, as is evident in the A2 distributions in Figure \ref{fig:transfer_factor_effect}, although this seems counterintuitive: the curves roll over into their steep decline at higher energies for smaller transfer factors $t$, but the curves also have much smaller values at those higher energy breaks. As $t$ decreases, the somewhat arbitrary definition of $E_{\hbox{heb}}$ for a smooth rollover has larger uncertainties for those cases with large bumps in the distribution (e.g., accelerator A2 in Figure \ref{fig:transfer_factor_effect}). 

\begin{figure}[ht]
   \includegraphics[width=\columnwidth]{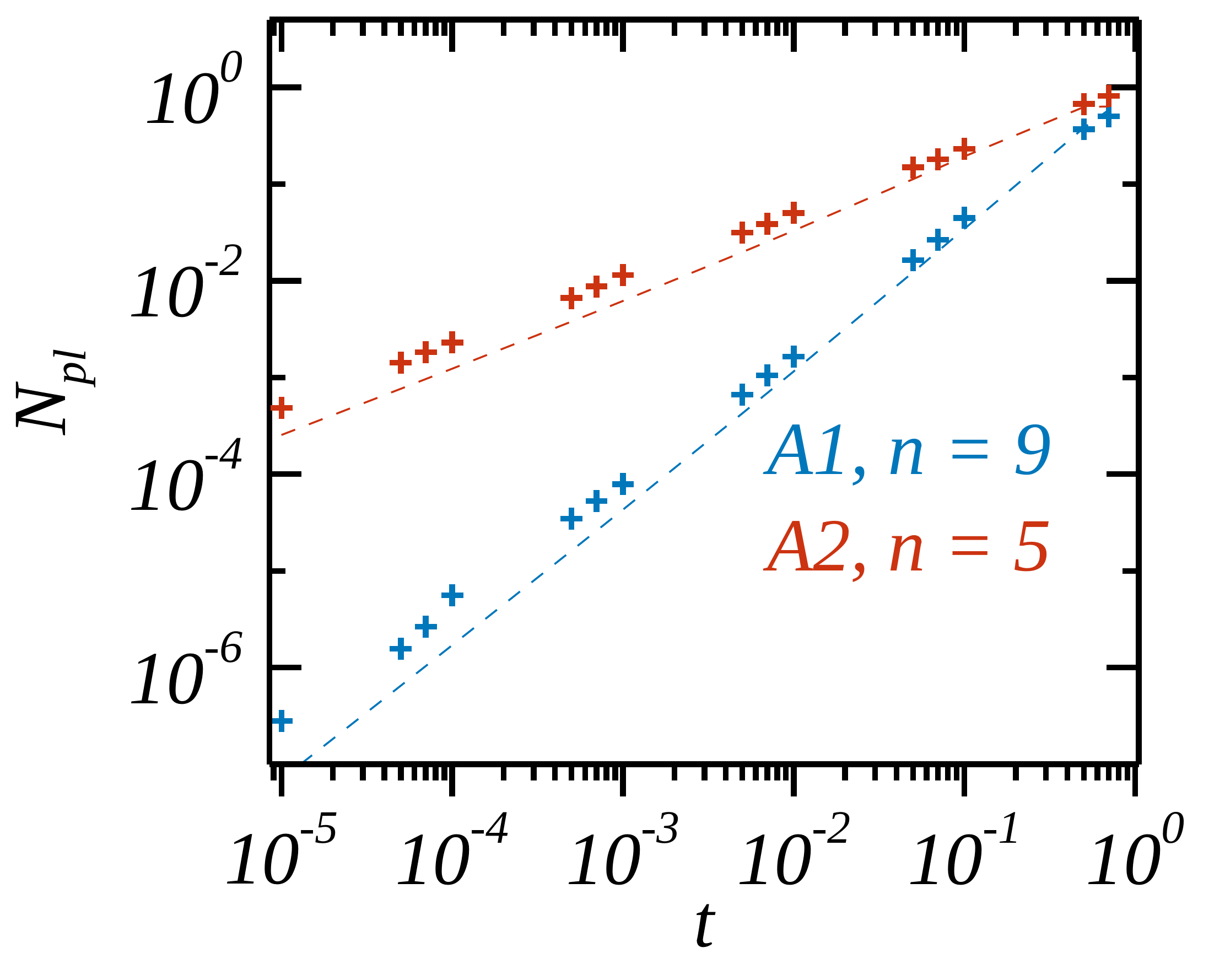}
   \caption{Number of particles in units of $N$ (total number of particles in each accelerator) in the power-law region as a function of the transfer factor $t$ for fixed $n$ (annotated). Accelerators A1 (A2) are represented by blue (red) crosses. The largest relative error is of a factor of $2.25$. Color-coded dashed lines are theoretically predicted values of $N_{\hbox{pl}}$ (Equation  \ref{eq:numbnonthanaly} in \S \ref{sec:analytical}, with the color-coded annotated $r$ shown in Figure\ \ref{fig:full_cycle_energy_gain} and with  $\alpha=0$).}
   \label{fig:numb_high_ener_part}
\end{figure}

We also calculated the fractional number of high-energy particles in the power-law region between the low- and high-energy breaks, $N_{\hbox{\hbox{pl}}}$. The results are shown in Figure \ref{fig:numb_high_ener_part} for A1 (blue) and A2 (red) as functions of the transfer factor $t$ for fixed $n$ (annotated). $N_{\hbox{pl}}$ closely follows a positive power-law trend versus $t$, showing how transferring more particles between accelerators yields more particles in the most energized region of the final distribution. The stronger accelerator, A2, has substantially more energized particles than the weaker accelerator, A1, especially at small transfer factors $t$. However, the number of particles is more sensitive to $t$ for A1 compared to A2, as indicated by the steeper slope of the blue curve in the figure. Augmenting the number of visited accelerators, $n$, in either case results in more particles in the power-law region of the spectrum as the high-energy break occurs at higher energies.

\begin{figure}[ht]
   \includegraphics[width=\columnwidth]{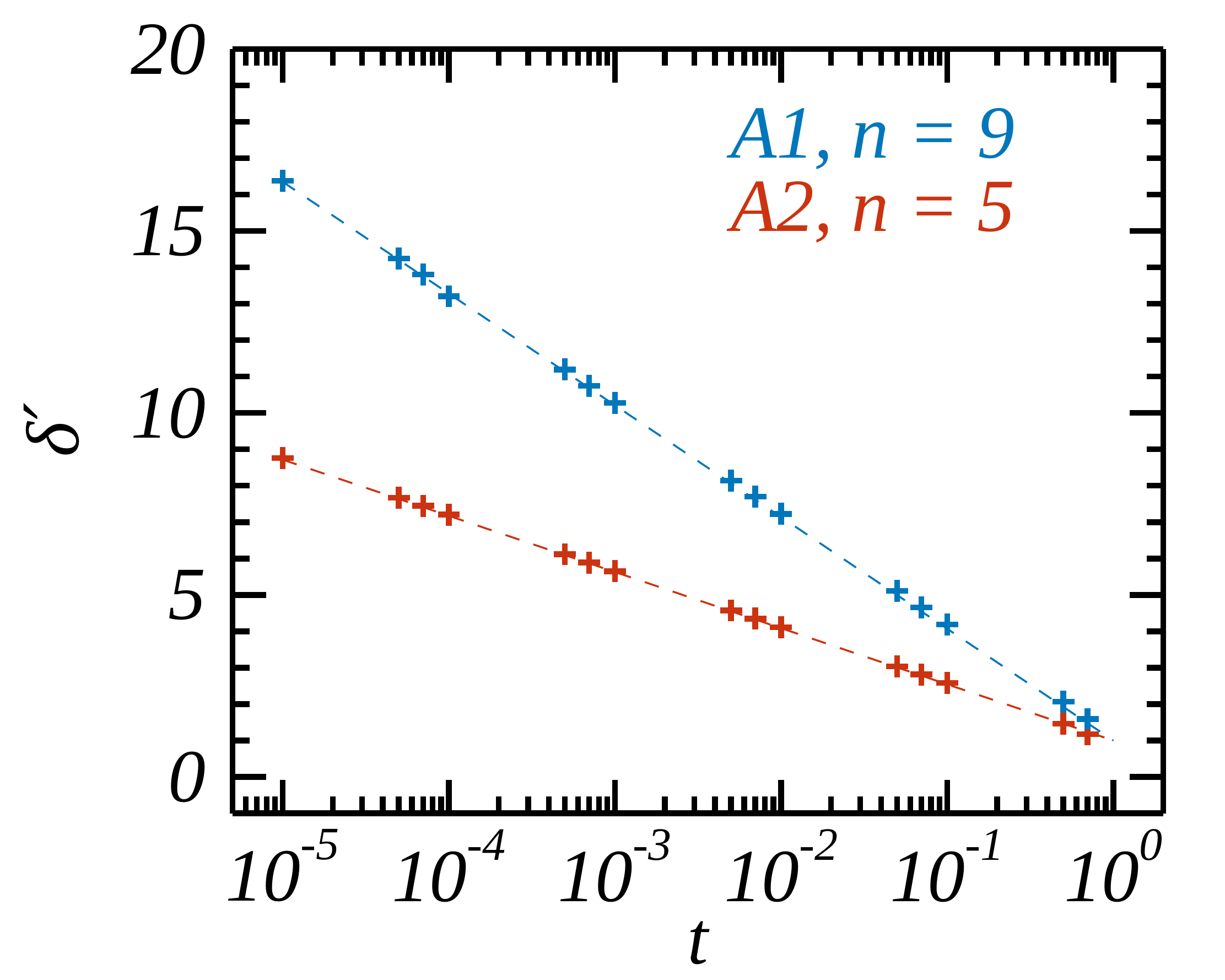}
   \caption{Spectral index as a function of $t$ for fixed $n$ (annotated). Fitted $\delta'$ for A1 (A2) are shown with blue (red) crosses. The average percent error is $<1$\%. Color-coded dashed lines are theoretically predicted values of $\delta'$ (Equation\ \ref{eq:most_important} in \S \ref{sec:analytical}, with the color-coded annotated $r$ shown in Figure\ \ref{fig:full_cycle_energy_gain} and with  $\alpha=0$). }
   \label{fig:efficiency_dependence}
\end{figure}

The fitted spectral indices $\delta'$ as function of $t$ for A1(blue) and A2 (red) are shown with crosses in Figure \ref{fig:efficiency_dependence}. The indices follow a logarithmically decreasing dependence, indicating increasingly hard spectra, as the transfer factor $t$ increases. The errors in the fitted spectral indices generally are less than $0.1\%$.  A1's spectral indices are larger (softer) than A2's because A1's energy gains in each cycle are smaller and, hence, change more slowly with $t$. The hardening of the spectrum for A2 at increasing values of $t$ is evident in Figure \ref{fig:transfer_factor_effect}. 
\section{Analytical Model}
\label{sec:analytical}

This section demonstrates that the key features of the numerical spectra from the previous section can be reproduced and analyzed with a fully analytical model with a simple assumption: particle acceleration is performed sequentially in accelerators with modest energy gains. This model emulates basic features of the final particle distribution, such as spectral index, energy breaks, bumps in the distribution, and other details of the energy distributions as functions of very few physical parameters. 

In this model, each accelerator evolves an initial particle distribution into another distribution by means of an \textit{unspecified} acceleration mechanism. We simplify the details of the mechanism by assuming that each cycle increases the average particle energy by a factor $r$, which we denote the ``efficiency'' of the accelerator. This results in a strictly exponential increase in the average energy, similar to that of the island acceleration mechanism shown in Figure \ref{fig:full_cycle_energy_gain}. Distributions are assumed to be summed over pitch angle, so only energy dependence is considered. As before, all accelerators are assumed to have the same average characteristics, specifically $t$, $r$, and $N$. (Definitions of the model parameters are summarized in Table \ref{tab:table2}.) 

Based on the distributions obtained for accelerator A2 (Figure \ref{fig:each_cycle}), we assume that each cycle results in a final population described by an analytical function that scales in a self-similar way from its initial population, with the average energy increasing by a factor $r$. Two simple, well-known such distributions are represented by the Maxwellian and exponential functions. These are special cases of a more general function of variables $R$ (proportional to the average energy of the distribution) and $\alpha$ ($\alpha = 0$ for exponential and $\alpha = 1/2$ for Maxwellian), all of which are listed in Table \ref{tab:table3}.  In the Maxwellian case, the thermodynamic temperature is well defined, and each cycle simply heats the particles from temperature $T$ to temperature $rT$. 
\begin{table}[ht]
  \begin{center}
    \caption{Model parameters.}
    \label{tab:table2}
    \begin{tabular}{|l|l|}
     \hline
      \textbf{Symbol} & \textbf{Definition} \\
      \hline
      \hline
        \tablespaceaddtwo
            $t$ & Fraction of particles transferred   \\
            & between accelerators  \\  [1ex]
        \hline
        \tablespaceaddtwo
            $r$ & Accelerator efficiency \\  
      \hline       
        \tablespaceaddtwo
            $n$ & Number of accelerators  \\
            & visited by particles \\   [1ex]
     \hline   
           \tablespaceaddtwo
            $N$ & Total number of particles  \\ 
            & in each accelerator \\ 
            &  (distribution function   \\
             &  normalization constant)
             \\
             [1ex]
        \hline
\end{tabular}
  \end{center}
\end{table}

Empirically, we found that the exponential function was a better fit for our simulated distributions from \S \ref{sec:results} than the Maxwellian. Exponential fits $e^{-\overline{E}/R}/R$ to A2 distributions are shown in Figure \ref{fig:each_cycle} as black dashed lines for each cycled distribution. Fitted $R$ values for A1 and A2, which are equal to $\left <\overline{E}\right>$ for exponential functions, are shown in Figure \ref{fig:full_cycle_energy_gain} with color-coded circles. The characteristic efficiency $r$ for each accelerator was found by fitting the derived values of $R$ as a function of each cycle $j$ with function $r^{j}$ (see Table \ref{tab:table3}), resulting in $r \simeq 2.117$ for A1 and $r \simeq 4.446$ for A2 (also annotated in Figure\ \ref{fig:full_cycle_energy_gain}). These values quantify in a simple way how much more efficient A2 is at accelerating particles than A1. 

\begin{table*}[ht]
  \begin{center}
    \caption{Distribution functions for the analytical model.}
    \label{tab:table3}
    \begin{tabular}{|c|c|c|c|} 
    \hline
         \textbf{Type } &   \textbf{Functional Form $f$} &   \textbf{Iterative Form $f_j$} & \textbf{Average Energy} \\
          &  & \textbf{ with} $R=r^{j}$ &  $\left<\mathpunct{\overline{E}}\right>_j = \dfrac{\left<E\right>_j}{k_{B} T}$\\
          [3ex]
           \hline
           \hline 
           \tablespaceadd
                General Form  &  \vspace{1ex} $   \dfrac{1}{\Gamma(\alpha+1)} \left(\dfrac{\overline{E}}{R}\right)^\alpha \dfrac{e^{- \overline{E}/R } }{R}    $ &  \vspace{1ex} $  \dfrac{1}{\Gamma(\alpha+1)} \left(\dfrac{\overline{E}}{r^{j}}\right)^\alpha \dfrac{e^{- \overline{E}/r^{j} } }{r^{j}}    $ & $ \dfrac{\Gamma(\alpha+2)}{\Gamma(\alpha+1)} R =  \dfrac{\Gamma(\alpha+2)}{\Gamma(\alpha+1)} r^{j}$  \\ [3ex]
            \hline 
            \tablespaceadd
                Exponential ($\alpha=0$) &  \vspace{3ex} $ \dfrac{ e^{-  \overline{E} /R}  }{R} $ &  \vspace{3ex} $  \dfrac{ e^{-  \overline{E} /r^j}  }{r^j} $ & $  R = r^{j}  $ \\ [3ex]
            \hline 
            \tablespaceadd
               Maxwellian ($\alpha=\frac{1}{2}$)  &  \vspace{1ex} $  \dfrac{2}{\sqrt{\pi}} \sqrt{\dfrac{\overline{E}}{R}} \dfrac{ e^{- \overline{E}/R } }{R}    $ &  \vspace{1ex} $  \dfrac{2}{\sqrt{\pi}} \sqrt{\dfrac{\overline{E}}{r^{j}}} \dfrac{e^{- \overline{E}/r^{j} } }{r^{j}}    $ &   $ \dfrac{3}{2} R = \dfrac{3}{2} r^{j}$  \\ [3ex]
       \hline
       \multicolumn{4}{l}{\footnotesize{$\Gamma$: complete Gamma function, where $\Gamma(x)=(x-1)!$, with $x \in \mathbb{R}$ ($\Gamma(1)=1$,$\Gamma(3/2)=\sqrt{\pi}/2$).}} \\
 \end{tabular}
  \end{center}
\end{table*}

For the sake of greater generality, however, we adopt the general analytical form from Table \ref{tab:table3} in the following calculations because Maxwellian distributions are assumed so widely in solar flare studies.  With these assumptions, we construct the final distribution of a population formed by sequential acceleration with efficiency $r$ and transfer factor $t$ by explicitly substituting the general form of the functions $f_j$ (Table \ref{tab:table3}) into the final distribution function $f_f^{(n)}$ (Equation \ref{eq:overlinedist1}): 
\begin{eqnarray}
\label{eq:dist_anal_form}
 f_f^{(n)}  =  \left(\dfrac{1-t}{t}\right)  \sum \limits_{j=1}^{n}  
 \dfrac{e^{- \overline{E}/r^{j} } }{\Gamma(\alpha+1)} \left(\dfrac{\overline{E}}{r^{j}}\right)^\alpha  \left(\dfrac{t}{r}\right)^j  \nonumber \\
  + \dfrac{e^{- \overline{E}/r^{n} } }{\Gamma(\alpha+1)}
 \left(\dfrac{\overline{E}}{r^{n}}\right)^\alpha  \left(\dfrac{t}{r}\right)^n,  
\end{eqnarray}
where $\Gamma(x)=(x-1)!$ is the complete Gamma function (see footnote on Table \ref{tab:table3}).

\begin{figure}[ht]
   \includegraphics[width=\columnwidth]{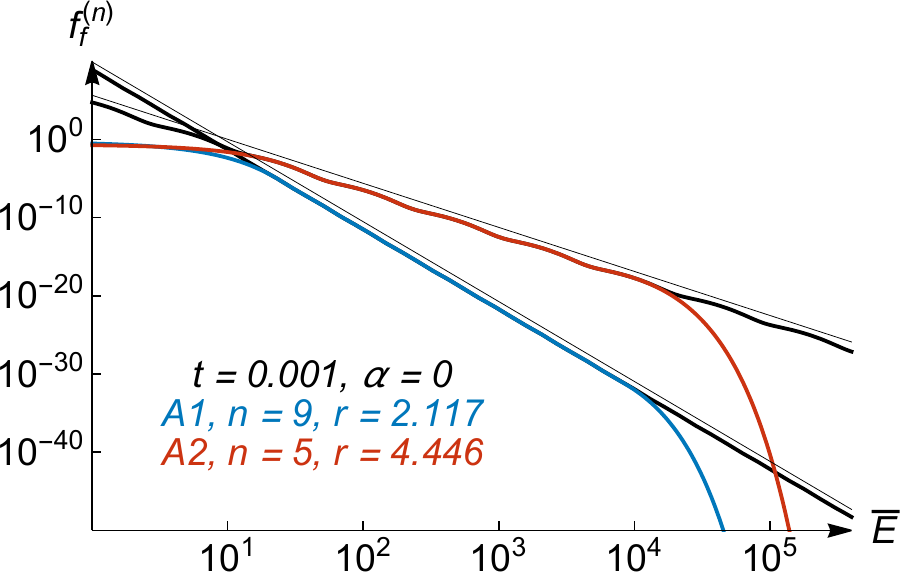}
   \caption{Final distributions as functions of $\overline{E}$ from Equation \ref{eq:dist_anal_form}, with color-coded annotated parameters.  $r$ values are fitted values for A1 and A2 (see Figure\ \ref{fig:full_cycle_energy_gain}). Thick black curves show Equation \ref{eq:dist_anal_form_per} evaluated for the finite sum $j \in [-50,+50]$ (Equation \ref{eq:dist_anal_form_7}); thin black curves show the power law $C E^{-\delta'}$ (shifted up slightly for clarity) using $C$ from Equation \ref{eq:c_const} and $\delta'$ from Equation \ref{eq:most_important}.
   }
   \label{fig:analyti_sol}
\end{figure}

The supplemental Wolfram Mathematica notebook ``Guidoni\_etal\_Suppl\_Math\_Notebook.nb'' provides a widget that plots $f_f^{(n)}$ (Equation\ \ref{eq:dist_anal_form}), where the user can explore the parameter space ($t,r,n,\alpha$).

As an intermediate check, we constructed distributions from Equation \ref{eq:dist_anal_form} using $t=0.001$, $\alpha = 0$, and the fitted values of $r$ for A1 and A2 and compared them to the simulated data in Figure \ref{fig:mirror_transit_index}. The analytical curves in that figure (dashed lines) have been shifted down for visual clarity because they overlap the simulated curves, corroborating our results. These curves are also plotted with blue and red solid lines in Figure\ \ref{fig:analyti_sol} to be compared to other curves presented in this section.

The average energy of $f_f^{(n)}$ is evaluated (using Table \ref{tab:table3} and Equation \ref{eq:dist_anal_form}) and expressed in the alternative forms 
\begin{eqnarray}
\label{eq:aveeneranaly}
 \left<\overline{E}\right>^{(n)} 
 & = & \dfrac{\Gamma(\alpha+2)}{\Gamma(\alpha+1)} 
 \left[\dfrac{r (1-t) -(tr)^n(r-1)}{1-tr}\right] \\
 & = & \dfrac{\Gamma(\alpha+2)}{\Gamma(\alpha+1)} 
 \left[r + (r-1)tr \dfrac{1-(tr)^{n-1}}{1-tr}\right]. \nonumber
\end{eqnarray}

In the limit $tr \ll 1$, only the leading $r$ term in the brackets above is important, and  $\left<\overline{E}\right>^{(n)} \approx \left<\overline{E}\right>^{(1)}$, the average energy after the first cycle. In this case, as explained in \S \ref{sec:results}, not much additional energy is gained subsequently by the system. This is illustrated in Figure \ref{fig:high_energy_break}, where $\left<\overline{E}\right>^{(n)}$ essentially is unchanged as more cycles are added beyond $n = 1$. 

We demonstrate that the middle-energy range of $f_f^{(n)}$ approximates a power law in $\overline{E}$ by writing Equation \ref{eq:dist_anal_form} in the form 
\begin{eqnarray}
\label{eq:dist_anal_form_2}
f_f^{(n)} = \left(\dfrac{1-t}{t}\right) \dfrac{{\overline{E}}^{-\delta'}}{\Gamma(\alpha+1)} g_f^{(n)},
\end{eqnarray}
where
\begin{eqnarray}
\label{eq:dist_anal_form_3}
g_f^{(n)}  = \sum \limits_{j=1}^{n}  
 e^{- \overline{E}/r^{j} } \left(\dfrac{\overline{E}}{r^{j}}\right)^\alpha  {\overline{E}}^{\delta'} \left(\dfrac{t}{r}\right)^j \nonumber \\
           + \left(\dfrac{t}{1-t}\right) e^{- \overline{E}/r^{n} } \left(\dfrac{\overline{E}}{r^{n}}\right)^\alpha  {\overline{E}}^{\delta'} \left(\dfrac{t}{r}\right)^n.
\end{eqnarray}
The above equation takes a simple form if we define the auxiliary variable $x$,
\begin{eqnarray}
\label{eq:dist_anal_vars}
x &\equiv& \log \overline{E},
\end{eqnarray}
and choose 
\begin{eqnarray}
\label{eq:most_important}
\delta' &=& 1-\dfrac{\log t}{\log r},
\end{eqnarray}
whence $t/r = r^{-\delta'}$. Note that $\delta' > 1$ because $t < 1$ and $r > 1$; furthermore, $\delta'$ is large if $tr \ll 1$.

We then obtain the expression
\begin{eqnarray}
\label{eq:dist_anal_form_5}
g_f^{(n)}(x) = \sum \limits_{j=1}^{n}  g \left( x - j \log r \right) \nonumber \\ + \left(\dfrac{t}{1-t}\right) g \left( x - n \log r \right) .
\end{eqnarray}
The function $g(x)$ is defined by 
\begin{eqnarray}
\label{eq:dist_anal_form_6}
g(x) &=& e^{-10^x + \left(\alpha+\delta' \right) \ln 10^x }.
\end{eqnarray}

\begin{figure}[ht]
   \includegraphics[width=\columnwidth]{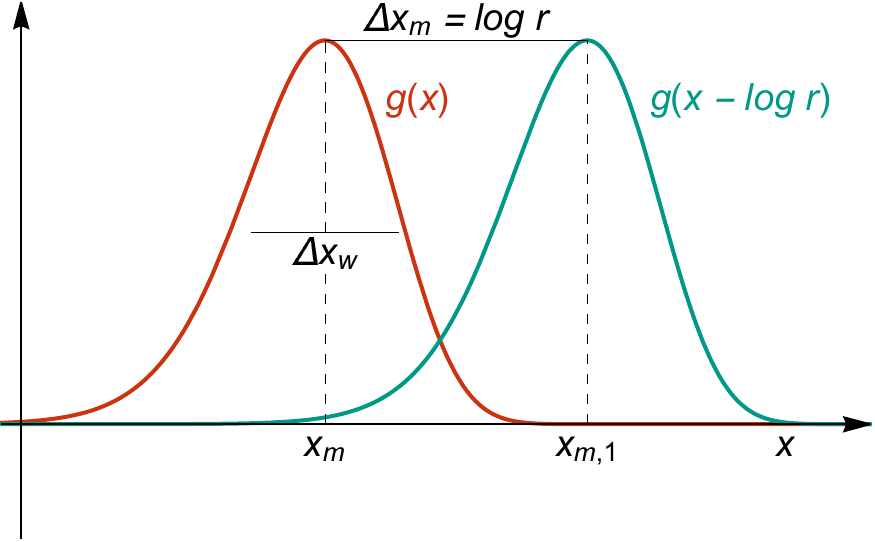}
   \caption{Sketch of the function $g(x)$ in Equation \ref{eq:dist_anal_form_6} (red) and its parameters, along with the next shifted pulse (green). For this example, $\alpha=0$, $r$ = 4.446 (fitted value for A2), and $t = 0.001$ (same $t$ as in Figure \ref{fig:mirror_transit_index}).}
   \label{fig:gfunction}
\end{figure}

Equation \ref{eq:dist_anal_form_5} is a sum of positive, identically shaped pulse-like functions $g(x)$ spaced at equal intervals $\log r$. Two examples of consecutive pulses, $g(x)$ and $g(x-\log r)$, are shown in Figure \ref{fig:gfunction}. $g(x)$ attains its maximum value at $x_m = \log \left( \alpha+\delta' \right)$ and decays in both directions from its peak, at the rate $\left( \alpha+\delta' \right) \ln 10 $ in the negative direction and at rate $-10^x \ln 10$ in the positive direction. Each $j$ term in the $g_f^{(n)}$ expansion, peaks at 
\begin{eqnarray}
\label{eq:dist_anal_max_2}
x_{m,j} &=& x_m + j \log r
\end{eqnarray}

\begin{figure}[ht]
   \includegraphics[width=\columnwidth]{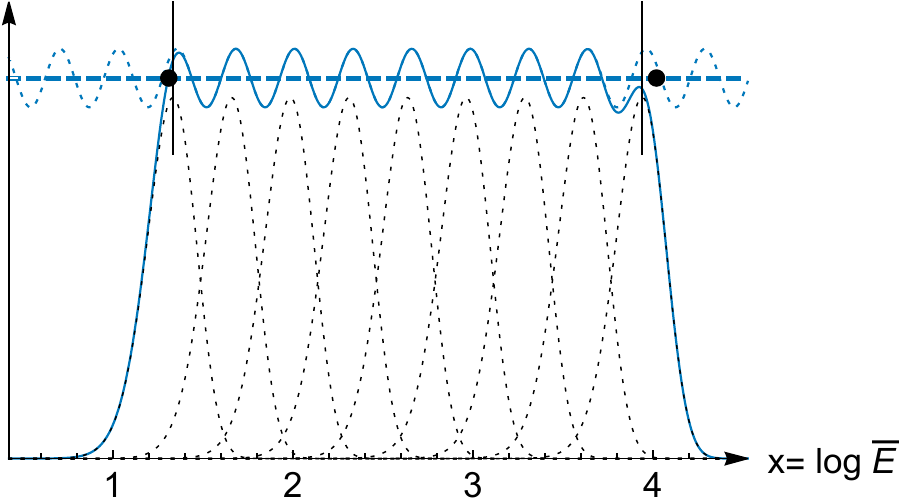}
   \caption{Sketch of the function $g_f^{(n)}(x)$ (Equation \ref{eq:dist_anal_form_5}, solid blue). For this example, $\alpha=0$, $r= 2.117$ (fitted value for A1), and $t = 0.001$ (same $t$ as in Figure \ref{fig:mirror_transit_index}). Each term of the sum in Equation \ref{eq:dist_anal_form_5} is drawn with a black dashed line (pulse-like curves). The left (right) vertical segment in black is located at the maximum of the first (last) pulse, $x_{m,1}$ ($x_{m,n}$), marking the approximate start (end) of the power-law-like region of $f_f^{(n)}$. Black circles indicate fitted $\log{E_{\hbox{leb}}}$ (left) and $\log{E_{\hbox{heb}}}$ (right) for A1 accelerators (\S \ref{sec:fitsection}, dashed vertical blue lines in Figure \ref{fig:mirror_transit_index}). The sinusoidal dashed blue curve shows Equation \ref{eq:dist_anal_form_7} evaluated for the finite sum $j \in [-10 n,+10 n]$ with $n=9$. The horizontal dashed blue line shows $\mathcal{F}_0$, the Fourier coefficient $k=0$ in Equation \ref{eq:dist_anal_form_8}.}
   \label{fig:gfnexpansion}
\end{figure}

The sum of the pulses in Equation \ref{eq:dist_anal_form_5} results in a function localized in the region between $x_{m,1}$ and $x_{m,n}$, which quickly decays to zero outside this interval. An example of $g_f^{(n)}$ is shown in Figure \ref{fig:gfnexpansion} (solid blue), where each term (pulse) of the sum in Equation \ref{eq:dist_anal_form_5} is shown with a dashed black line. $g_f^{(n)}$ oscillates about an approximate constant value (e.g., horizontal blue dashed line in Figure \ref{fig:gfnexpansion}) in the region between $x_{m,1}$ and $x_{m,n}$ (vertical black segments in Figure \ref{fig:gfnexpansion}). The supplemental Wolfram Mathematica notebook ``Guidoni\_etal\_Suppl\_Math\_Notebook.nb'' provides a widget that plots $g_f^{(n)}$ as in Figure \ref{fig:gfnexpansion}, where the user can explore the parameter space ($t,r,n,\alpha$). 

The oscillation amplitude of $g_f^{(n)}$ decreases (increases) as the overlap between its pulses increases (decreases) because the weight of each pulse (cycle) on the overall curve decreases (increases). It is straightforward to show from Equation \ref{eq:dist_anal_form_6} that, at the location of its maximum, $x_m = \log \left( \alpha+\delta' \right)$, the maximum value of $g(x)$ and its second derivative are, respectively, 
\begin{eqnarray}
\label{eq:dist_anal_max}
g(x_m) &=&  \left( \dfrac{\alpha+\delta'}{e} \right)^{\alpha+\delta'}\hbox{ and} \\
g''(x_m) &=& - \left( \alpha+\delta' \right) g(x_m)(\ln 10)^2. \nonumber
\end{eqnarray}
The above second derivative $g''\left( x_m \right)$ shows that the pulse typically is localized about $x = x_m$. The pulse width is then $\Delta x_w \simeq  2 \sqrt{-g(x_m)/g''(x_m)} = 2 \left( \alpha+\delta' \right)^{-1/2}/\ln 10$. Therefore, for a fixed pulse peak separation (fixed $r$), the width of a pulse (and consequently its overlap with neighboring pulses) increases with $t$ ($\delta'$ decreases with $t$, see Equation\ \ref{eq:most_important}). In Figure \ref{fig:transfer_factor_effect}, for example, the oscillations of the distributions decrease in amplitude with increasing $t$ because the width of the pulses that compose $g_f^{(n)}$ increase with that parameter.  

For a large portion of the ($t,r,n$) parameter space, therefore, $f_f^{(n)}$ can be approximated by a power law with spectral index $\delta'$ modulated by the $g_f^{(n)}$ oscillations (see Equation \ref{eq:dist_anal_form_2}). Figure \ref{fig:efficiency_dependence} shows the predicted $\delta'$ values for A1 and A2 from Equation \ref{eq:most_important} (dashed), which agree closely with the fitted values determined in \S \ref{sec:fitsection} (crosses). 

Converting $x_{m,1}$ to energy, we obtain for the approximate location of the low-energy break 
\begin{eqnarray}
\label{eq:dist_anal_max_6}
\overline{E}_{\hbox{leb}} &\sim& 10^{x_{m,1}} = \left( \alpha+\delta' \right) r = \left(\alpha + 1 \right) r - \dfrac{r \log t}{\log r}.
\end{eqnarray}
This result is consistent with accelerators A1 and A2 having $\overline{E}_{\hbox{leb}}$ with an approximate logarithmic dependence on $t$, as shown in Figure\ \ref{fig:lowenergybreak} (where Equation \ref{eq:dist_anal_max_6} is shown with color-coded dashed lines for $r=2.117$ and $r=4.456$). The nearly identical slopes for A1 and A2, despite their very different efficiencies --- $r \approx 2 $ (A1) and  $r \approx 4$ (A2) --- are a consequence of the similar ratios: $r/\log r \approx 2/\log 2 \approx 4/\log 4$. For comparison, the corresponding fitted $ \overline{E}_{\hbox{leb}}$ in log space from Section\ \ref{sec:fitsection} is shown with the left black circle in Figure\ \ref{fig:gfnexpansion}. 

Similarly, the high-energy break of the power law occurs near the last ($j=n$) peak,
\begin{eqnarray}
\label{eq:dist_anal_max_9}
\overline{E}_{\hbox{heb}} &\approx& 10^{x_{m,n}} \approx \left(\alpha + 1 \right) r^n - \dfrac{r^n \log t}{\log r}.
\end{eqnarray}
This result is consistent with A1 and A2 having similar high-energy breaks $\overline{E}_{\hbox{heb}}$, as illustrated by Fig.\ \ref{fig:highenergybreak} (where Equation \ref{eq:dist_anal_max_9} is shown with color-coded dashed lines). Essentially, the difference in the number of visited accelerators compensates for the difference in efficiencies.  The high-energy breaks for A1 and A2 differ somewhat more than their low-energy breaks and the variations with $\log t$ deviate rather more from the linear relationship indicated by Equation \ref{eq:dist_anal_max_9}. The fitted $\overline{E}_{\hbox{heb}}$ in log space from Section\ \ref{sec:fitsection} is marked by the right black circle in Figure\ \ref{fig:gfnexpansion}.

The extent of the power law is then $\overline{E}_{\hbox{heb}}/\overline{E}_{\hbox{leb}} \simeq 10^{(n-1) \log r}$. The smaller (larger) the efficiencies of the accelerators, the larger (smaller) the number of cycles required to develop a power law of a given range. Figures \ref{fig:mirror_transit_index} and \ref{fig:analyti_sol} demonstrate that A1 accelerators need 9 cycles to achieve a similar power-law range as 5 cycles of A2 accelerators ($8 \log{2.12} \simeq 4 \log{4.45} \simeq 2.6 $). Additional cycles with a given efficiency also extend the region of the power law.

To determine the approximate constant value about which $g_f^{(n)}$ oscillates, we note that in that region the contributions of terms in Equation \ref{eq:dist_anal_form_5} that peak toward the ends of the power-law energy range become increasingly small near the center of the range; in particular, if $t$ is small, the pulses are narrow and the last term in the sum is negligible. As an approximation, therefore, we extend the summation in Equation \ref{eq:dist_anal_form_5} to include all integers $j<1$ and $j>n$:
\begin{eqnarray}
\label{eq:dist_anal_form_7}
g_f^{(n)} &\approx& g_f^{\infty} \equiv \sum \limits_{j=-\infty}^{+\infty}  g \left( x - j \log r \right).
\end{eqnarray}
An example of Equation \ref{eq:dist_anal_form_7} evaluated for the finite sum $j \in [-10 n,+10 n]$ with $n=9$ is shown with the sinusoidal blue dash line in Figure \ref{fig:gfnexpansion}. Using more terms does not change the results at the resolution of the graph. The approximate form $g_f^{\infty}$ in Equation \ref{eq:dist_anal_form_7} is explicitly periodic in $x$ with period $\log r$. Hence, it can be expressed as a Fourier series 
\begin{eqnarray}
\label{eq:dist_anal_form_8}
g_f^{\infty} = \sum \limits_{k=-\infty}^{+\infty}  e^{i 2 \pi k x / \log r} \mathcal{F}_k,
\end{eqnarray}
where $\mathcal{F}_k$ is the Fourier coefficient of mode $k$ in the space $x / \log r \in [-1/2,1/2]$. $g_f^{\infty}$ oscillates about $\mathcal{F}_0$, the value of the Fourier coefficient for $k=0$. Figure\ \ref{fig:gfnexpansion} shows $\mathcal{F}_0$ for A1 with a horizontal blue dashed line. The remaining coefficients for $k \ne 0$ are the amplitudes of oscillatory contributions to the full distribution that cause the latter to deviate from the strict power law. 

From Equation \ref{eq:dist_anal_form_2}, in the power-law region 
\begin{eqnarray}
\label{eq:dist_anal_form_per}
f_f^{(n)} \approx \left(\dfrac{1-t}{t}\right) \dfrac{{\overline{E}}^{-\delta'}}{\Gamma(\alpha+1)} g_f^{\infty}.
\end{eqnarray}
Thick black lines in Figure\ \ref{fig:analyti_sol} show Equation\ \ref{eq:dist_anal_form_per} solved with $g_f^\infty$ from Equation \ref{eq:dist_anal_form_7}, evaluated for the finite sum $j \in [-50,50]$. Using more terms does not change the results at the resolution of the graph. For both cases shown in the figure, the results very closely overlay the exact (blue and red) curves within the relevant power-law ranges and extend them smoothly to energies beyond both the low- and high-energy breaks. 

\begin{figure}[ht]
   \includegraphics[width=\columnwidth]{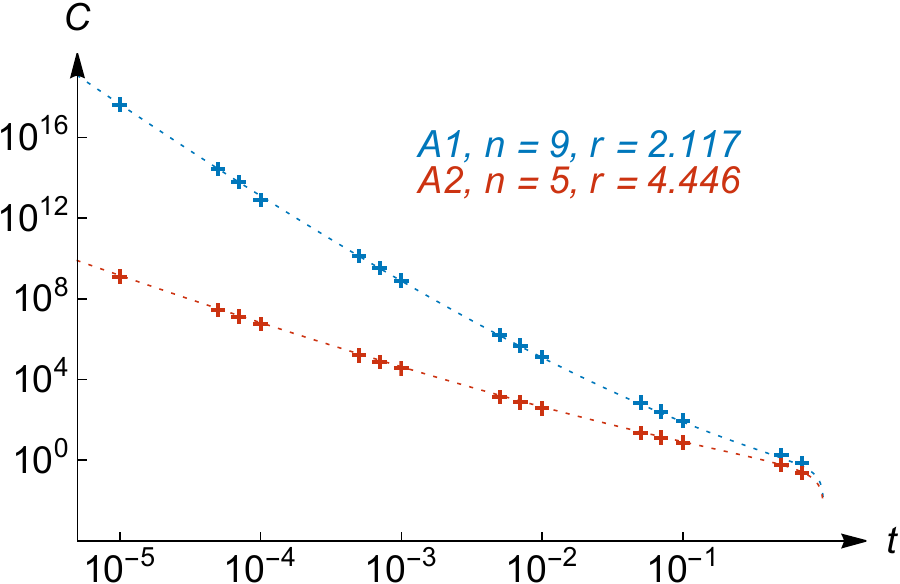}
   \caption{Analytical values of normalization constant $C$ (Equation \ref{eq:c_const}, dashed lines) for the color-coded annotated parameters. $r$ values are fitted values for A1 and A2 (see Figure\ \ref{fig:full_cycle_energy_gain}). Crosses show fitted $C$ for A1 (blue) and A2 (red) from fitting power laws to the distributions in \S \ref{sec:fitsection}, as described in the Appendix.}
   \label{fig:c_intercept.eps}
\end{figure}

The normalization constant of the power law is 
\begin{eqnarray} 
    C \approx  \left(\dfrac{1-t}{t}\right) \dfrac{\mathcal{F}_0}{\Gamma(\alpha+1)},
    \label{eq:c_const}
\end{eqnarray}
$C$ is displayed in Figure \ref{fig:c_intercept.eps}. Analytical values (Equation\ \ref{eq:c_const}, dashed) coincide with the fitted values (crosses) for accelerators A1 (blue) and type A2 (red) from Section\ \ref{sec:fitsection}, determined with the method described in the Appendix. The thin black curves in Figure\ \ref{fig:analyti_sol} show the analytical power-law $C \overline{E}^{-\delta'}$ (shifted up slightly for clarity) with $C$ from Equation \ref{eq:c_const} and $\delta'$ from Equation\ \ref{eq:most_important}. 

We estimated the number of particles in the power-law region by integrating
\begin{eqnarray}
 \label{eq:numbnonthanaly}
 N_{\hbox{pl}} & = & \int_{\overline{E}_{\hbox{leb}}}^{\overline{E}_{\hbox{heb}}}  C \overline{E}^{-\delta'} d\overline{E} \nonumber \\
 & = &  \left(\dfrac{1-t}{t}\right) \dfrac{\mathcal{F}_0 \left( \overline{E}_{\hbox{heb}}^{1-\delta'} - \overline{E}_{\hbox{leb}}^{1-\delta'}  \right)}{\Gamma(\alpha+1) (1-\delta')} 
\end{eqnarray}
Figure\ \ref{fig:numb_high_ener_part} compares the analytical (dashed) and numerical (crosses) values of $N_{\hbox{pl}}$. The analytical calculations underestimate the number of nonthermal particles because the analytical approximations for
$\overline{E}_{\hbox{leb}}$ (Equation\ \ref{eq:dist_anal_max_6}) and $\overline{E}_{\hbox{heb}}$ (Equation\ \ref{eq:dist_anal_max_9}) are larger and  smaller, respectively, than the fitted values (crosses in Figure\ \ref{fig:lowenergybreak} and \ref{fig:highenergybreak}). Almost identical results are obtained by integrating the distribution function in Equation\ \ref{eq:dist_anal_form}. Analytical Equations\ \ref{eq:cummul1}, \ref{eq:cummul2}, and \ref{eq:logdiffcummulanalyt} in the Appendix can be used to estimate the differences in $N_{\hbox{pl}}$ between a power law and the distribution function in Equation\ \ref{eq:dist_anal_form}. 
In summary, in this section we have shown that, for a large range of the ($t,r,n$) parameter space, sequentially accelerated-particle distributions have a range in energy where they can be approximated by a power law with spectral index $\delta'$ (Equation\ \ref{eq:most_important}). In addition, key features of the power law, such as energy breaks and number of nonthermal particles, can be estimated analytically and easily interpreted from few physical parameters. 

\section{Discussion}
 
We have investigated the acceleration of particles in the flaring solar corona by sequences of magnetic islands that form, contract, and are transported within the flare current sheet. Numerous islands populate the sheet, as has been shown by many eruptive-flare simulations, and by high-resolution, high-cadence observations of the Sun. In our previous study \citep{Guidoni_2016}, we analyzed the evolution of a few of these islands and their enclosed flux surfaces to determine their efficacy at accelerating particles. We found a maximum energy multiplication factor $\mathcal{E}_{max} \approx 4$ for the cases examined. This is significant, but it is not nearly sufficient to explain the high energies and power-law distributions of the electrons that generate hard X-rays in flares.

Consequently, in this paper we have investigated the effect of accelerating particles through multiple islands.
With an energy gain $\mathcal{E} = 4$ in each island, particles must visit only a few islands to increase their energies by orders of magnitude. For example, $n = 5$ such accelerators increase the energies of some particles by a cumulative factor $\mathcal{E}_{tot} \approx 1000$. We constructed sequences of distribution functions by assuming that a fraction $t$ of the particles accelerated in one island are transferred to the next island to receive another energy boost by a factor $r$. 

The distribution of ambient nonaccelerated particles at each stage is assumed to be an isotropic Maxwellian. For the island acceleration process studied here, the 
distribution of accelerated particles 
becomes increasingly anisotropic at each stage in the sequence. The degree of anisotropy depends upon the relative roles of betatron and Fermi acceleration in the contracting island, i.e., on the detailed changes in the island's size and shape as it traverses the flare current sheet. 

For our analysis, we did not separate mirroring from transiting populations as particles jump among accelerators because it is not clear how to characterize a change in pitch angle as particles move between accelerators. The total population (mirroring and transiting) was considered for the calculation of final spectra. 

We showed that the fitted spectra of the resulting energy distribution functions consist of a smooth, flat region at low energies, an approximately power-law region at intermediate energies, and a region with a sharply decreasing profile at high energies. The three regions are separated by low- and high-energy breaks. The power-law-like region presents some small bumps due to each acceleration cycle. For our simple model, we have assumed that particles are accelerated in a bath of accelerators whose properties can be described by averaged quantities. On the Sun, it is likely that this process will occur in multiple accelerators with different populations and values of the key parameters. The effect of inhomogeneous accelerators on the electron and photon spectra needs to be investigated.

We found that increasing the number $n$ of visited accelerators shifts the high-energy break to ever-higher energy, as expected, but it does not significantly change the spectral index $\delta'$ of the power-law region. In contrast, $\delta'$ depends sensitively upon the efficiency $r$ of the accelerators: larger $r$ broadens the distribution of each cycle more effectively than smaller $r$, so the index decreases and the spectrum becomes harder as $r$ increases. This is illustrated by the contrast between the distributions obtained with accelerators A1 and A2. Similarly, larger $t$ also broadens the distribution more effectively than smaller $t$, so that as with $r$, the index decreases and the spectrum becomes harder as $t$ increases.

To gain further insight into the results, we explored a simplified analytical model that emulates the average energy-amplification effect of the multiple-island acceleration mechanism while ignoring the effects on the isotropy of the distribution function. We found an analytical expression for the spectral index, $\delta' = 1 - (\log t) / (\log r)$, that replicates not only the qualitative features of our numerical results for the multiple-island model, but also the quantitative values of the index predicted by our numerical model. The analytic expression shows explicitly how changes in the transfer factor $t$ and the efficiency $r$ modify the index of the central power-law region of the energy spectrum. 

Our results also can be used to determine the transfer factor $t$ required to produce a measured spectral index $\delta'$, given an input efficiency $r$: $t = r^{1-\delta'}$. For an efficiency $r=4$ (our accelerator A2) and index $\delta'=5$, for example, $t = 4\times10^{-3}$. This is a tiny fraction of the particles resident in any island, but it is sufficient to produce a power law in the range typically inferred from solar flare observations. The required transfer factor $t$ depends strongly upon the efficiency $r$, however. For $r=2$ (our accelerator A1), as an example, $t = 6\times10^{-2}$, more than an order of magnitude greater than for the first case. However, we expect that efficiencies larger than $r=4$ might result for islands formed in flare current sheets with different parameters than those in our original simulated eruptive flare/CME (a more compact active-region source, higher field strengths, lower plasma $\beta$, etc.). If so, the necessary transfer factor $t$ would be smaller for the same index $\delta'$, or the index would be smaller for the same transfer factor. 

For simplicity, we assumed that $t$ is independent of energy in both the detailed modeling of the multiple-island mechanism and the streamlined analytical model. Our aim was to avoid artificially skewing the results toward producing power laws by supposing that the transfer of high-energy particles is more probable than that of low-energy particles. Because the high-energy particles actually are responsible for the power-law distribution, however, it seems likely that the transfer factor at the high-energy end of the spectrum ultimately determines the effective value of the transfer factor. In any case, a quantitative determination of $t$, via test-particle simulations or transport theory or some other means, would be invaluable, but is well beyond the scope of the present investigation. 

Also for simplicity, we further assumed that both $t$ and $r$ were the same throughout the sequence, as the particles were accelerated from one island to the next. Changes in the temperature $T$ of the bulk distribution over the lifetime of an island were ignored, as well. In the strongly time-varying environment of a flaring current sheet, all of these assumptions oversimplify the actual coronal evolution but enable us to make analytical progress and to interpret the results readily. However, the analytical model shows that the spectral index varies only logarithmically with the parameters $r$ and $t$. This weak dependence moderates the influence of relatively small -- factor-of-two or so -- variations in the parameters from time to time, or from point to point, within a single flare current sheet, or even from the current sheet in one flare to that in another. The ranges in the parameters $r$ and $t$ that are relevant to solar flares might be sufficiently limited to yield only a relatively narrow range of expected spectral indices $\delta'$. 

The  spectra of our analytical model can be easily used as injection populations in codes that model the transport of flare-accelerated particles from the top of flare arcades to their eventual thermalization at the solar surface \citep[e.g.,][]{Allred_2020}. The small number of parameters of our model simplifies the parameter-space exploration of the injection population when comparing the output of these codes with observed photon spectra. Determining $r$, $t$, and $n$ in this way provides average physical conditions of the acceleration region.

The hardest spectrum, i.e., the smallest value of the spectral index $\delta'$, is determined by the largest attainable values of $r$ and $t$ in combination. The highest energy that can be attained by a significant population of accelerated particles then is determined by $n$, the number of islands that a particle visits before it leaves the acceleration region. Assuming that thermal particles in the initial distribution are efficiently accelerated, the final distribution of particles is expected to extend over an energy range from $\overline{E} \approx 1$ to $\overline{E} \sim r^n$. The number of particles in the distribution falls by a factor $r^n$ as the number of particles is conserved during the acceleration process. For large transfer factors $t \lesssim 1$, these limits roughly define the extent of the power-law region of the distribution function. For smaller transfer factors $t \ll 1$, however, the power-law region shifts toward higher energies on both the low- and high-energy sides. The number of particles in the distribution declines steeply as the energy breaks shift. Hence, although the power-law region continues to span a large range in energy, it contains an increasingly small fraction of the particles as the transfer factor $t$ decreases. 

Altogether, our results suggest that particle acceleration during the contraction of multiple magnetic islands in current sheets may produce the high-energy particles that emit observed hard X-rays and microwaves in solar flares. Given a characteristic energy-amplification factor $r$ within single islands in the sheet, ultimately, accelerating many particles to high energies requires a significant fraction, $t$, of the particles to be transferred from one island to the next in the sequence, and for the particles to visit a sufficient number of islands, $n$, to achieve the needed energies. Our MHD simulations of eruptive flares have yielded initial values of $r \lesssim 5$ by exploring a limited parameter space that should be extended to include more compact flare source regions with higher magnetic-field strengths. Such simulations also could be used to determine the achievable values of the transfer factor $t$ and the number of visited accelerators $n$, by coupling the MHD model with a test-particle tracking model. This ambitious goal must be left to future investigations. 

Additional effects beyond the purview of MHD and test-particle tracking are important in a fully rigorous treatment of the problem of flare particle acceleration in coronal current sheets. First, we find the particle distributions that result from the process to be highly anisotropic. In a fully self-consistent kinetic calculation using PIC methods or the Vlasov-Maxwell equations, such particle distributions could initiate microinstabilities that induce electromagnetic field fluctuations. These fluctuations, in turn, would scatter the charged particles, altering the distribution of particle energies and angles from those calculated here. We point out that such effects could become important for \emph{any} model of flare particle acceleration that generates anisotropic distributions; this outcome is not limited to our simple model based on adiabatic invariants of the particle motion. 

Second, as in any test-particle calculation, there is no back reaction from the accelerated flare particles to the bulk plasma and magnetic field. In addition to inducing electromagnetic fluctuations, as just mentioned, the energized particles will exert their own thermal- and kinetic-pressure forces on the bulk plasma, carry electric currents, and drain energy from the magnetic field. All of these effects would modify the evolution of the system away from any elementary MHD description that does not account for them. This outcome, also, is not limited specifically to our model, and it could substantially alter the calculated particle distributions from the feedback-free case. 

These very challenging issues are being addressed by recent model advances developing from multiple perspectives. If kinetic-scale electric fields are not essential to the evolution of the system, as has been suggested by analyses of PIC simulations of particle acceleration by magnetic islands, one can apply a nonlinearly coupled, hybrid fluid/particle model suitable for collisional plasmas \citep{Drake_2019,Arnold_2019,Arnold_2021}. If the plasma is collisionless and turbulent, however, as is the case in the solar wind, guiding-center kinetic transport theory can be used to develop reduced prescriptions, including focused-transport theory and Parker transport equations, that describe the acceleration of particles by contracting and merging interplanetary flux ropes  \citep{Zank_2014,le_Roux_2015,le_Roux_2018,Zhao_2018,Adhikari_2019}. All of these developments seek to bridge the immense gulf between the governing macroscopic and microscopic scales at the Sun and in the heliosphere, and, at least in part, to explain the origin of high-energy particles in the solar system.

\acknowledgments

S.E.G was supported by NASA Grant 80NSSC 21K0817. J.T.K.\ and C.R.D.\ were supported by NASA's H-SR and H-ISFM programs. We are grateful to Meriem Alaoui Abdallaoui for useful discussions.

\appendix

\section{Spectral Fitting Method}
\label{sec:fitmethod}

Here we describe the automatic curve-fitting procedure used to estimate $\overline{E}_{\hbox{leb}}$, $\overline{E}_{\hbox{heb}}$, power-law parameters $C$ and $\delta'$, and their uncertainties, which requires minimal human intervention. 

The following procedural steps are performed with a (preferably) large number of iterations $m$, in each of which a small percentage $p$ of randomly selected points (3 points for the results in this paper) is withheld from the distribution to validate the model. For each resampled subset of points:

\begin{enumerate}
    \item \textbf{Manually initialize the high-energy break of the distribution, $\overline{E}_{\hbox{heb}}$}: visually determine an approximate value for $E_{\hbox{heb}}$. This is the only manual step of the whole procedure and the selected value does not have to be very accurate.
  
    \item \textbf{Fit a power law in the energy range $(\overline{E}_i, \overline{E}_{\hbox{heb}})$, for all $\overline{E}_i$ below $\overline{E}_{\hbox{heb}}$}: for each $\overline{E}_i$, perform a linear fit to determine $C$ and $\delta'$. 
  
    \item \textbf{Estimate the low-energy break $\overline{E}_{\hbox{leb}}$}: for each fitted power law, perform a Kolmogorov-Smirnov goodness-of-fit statistical test by computing the maximum of the absolute value of the difference between the empirical and theoretical complementary cumulative functions for each $\overline{E}_i$ \citep{Clauset_2009,Virkar_2014}. 
    The complementary cumulative functions are computed between energies $\overline{E}_{i}<\overline{E}_{d}<E_{\hbox{heb}}$ and $0<\overline{E}_{i}<E_{\hbox{heb}}$ as follows
        \begin{eqnarray}
             \label{eq:cummul1fit}
             F^{(p)}(\overline{E}_i,\overline{E}_d)  & = &   \int_{\overline{E}_{i}}^{\overline{E}_{d}}  C \overline{E}^{-\delta'} d\overline{E} \\ 
              \label{eq:cummul2fit}
              F^{(f)}(\overline{E}_i,\overline{E}_d)  & = &
              \int_{\overline{E}_{i}}^{\overline{E}_{d}}  f_f^{(n)} d\overline{E}.
        \end{eqnarray}
   Define a function $F(\overline{E}_i)$ as the maximum of the absolute value of the difference between the above complementary cumulative distributions
         \begin{eqnarray}
             \label{eq:diffcummul}
             F(\overline{E}_i) & = & \hbox{Max}\left|F^{(p)}(\overline{E}_i,\overline{E}_d) - F^{(f)}(\overline{E}_i,\overline{E}_d)\right|.
        \end{eqnarray}
    $F(\overline{E}_i)$ may have several local minima due to the bumps in the distributions of each cycle. $\overline{E}_{\hbox{leb}}$ is chosen as the energy $\overline{E}_i$ that corresponds to the first local minimum (lowest energy) of $F(\overline{E}_i)$. 

    \item \textbf{Estimate the high-energy break $\overline{E}_{\hbox{heb}}$}: Due to the much smaller values of the particle distribution function at high energies, we chose a method where small fluctuations have less impact. We computed the difference in area under the logarithm of the empirical and theoretical distribution functions. These areas in logarithmic space $x = \log \overline{E}$ are 
         \begin{eqnarray}
             \label{eq:logcummul1}
            logF^{(p)}(\overline{E}_{d}) & = &  \int_{x_{\hbox{leb}}}^{x_{d}} \log\left(C \overline{E}^{-\delta'}\right)  dx =  \int_{\overline{E}_{\hbox{leb}}}^{\overline{E}_{d}} \left[ \log (C) - \delta' \log[\overline{E}]  \right] \frac{d\overline{E}}{\overline{E} \ln(10)} \\ 
              \label{eq:logcummul2}
            logF^{(f)}(\overline{E}_{d})  & = &  \int_{x_{\hbox{leb}}}^{x_{d}} \log[f_f^{(n)}] dx =
              \int_{\overline{E}_{\hbox{leb}}}^{\overline{E}_{d}} \log[f_f^{(n)}] \frac{d\overline{E}}{\overline{E} \ln(10)}.
        \end{eqnarray}
     Then, we define a function $logF$ as the absolute value of the difference between the areas defined above 
     \begin{eqnarray}
             \label{eq:logdiffcummul}
             logF(\overline{E}_{d}) & = & \left|logF^{(p)}(\overline{E}_{d}) - logF^{(f)}(\overline{E}_{d})\right|.
     \end{eqnarray}
     $\overline{E}_{\hbox{heb}}$ is chosen as the $\overline{E}_d$ that corresponds to the last local minimum (highest energy) of $logF(\overline{E}_d)$. It is worth noticing that the bump at $t = 0.05$ for the A1 high-energy break in Figure \ref{fig:highenergybreak} remains after changing the seed of the random generator of the fitting method.
   
    \item \textbf{Determine $C$ and $\delta'$}: Perform a final linear fit in the energy range $(\overline{E}_{\hbox{leb}}, \overline{E}_{\hbox{heb}})$, to determine $C$ and $\delta'$. 
\end{enumerate}

The final $\overline{E}_{\hbox{leb}}$, $\overline{E}_{\hbox{heb}}$, $C$, and $\delta'$ and their uncertainties are calculated as the mean and the standard deviation over the $m$ iterations of the quantities estimated in the above procedure. To find local minima in noisy data, we smoothed differences in empirical and theoretical data with a box of width = 11 points. 

\section{Analytical Procedure}

Here we provide an alternative method for estimating estimate energy breaks for the analytical distribution functions in \S \ref{sec:analytical} using a method similar to the one described in \S \ref{sec:fitmethod}.

\begin{enumerate}
    \item Use Equation \ref{eq:dist_anal_max_9} as the initial approximation for $\overline{E}_{\hbox{heb}}$. 
    \item Use $C$ and $\delta'$ from Equations \ref{eq:c_const} and \ref{eq:most_important}, respectively. 
   
    \item The complementary cumulative functions (Equations\ \ref{eq:cummul1fit} and \ref{eq:cummul2fit}) in this case are
        \begin{eqnarray}
             \label{eq:cummul1}
             F^{(p)}(\overline{E}_i,\overline{E}_d)  & = &  \left(\dfrac{1-t}{t}\right) \dfrac{\mathcal{F}_0}{\Gamma(\alpha+1) (1-\delta')} \left( \overline{E}_{d}^{1-\delta'} - \overline{E}_{i}^{1-\delta'}  \right) \\ 
              \label{eq:cummul2}
              F^{(f)}(\overline{E}_i,\overline{E}_d)  & = &  \left. \dfrac{-1}{\Gamma(\alpha+1)} \left[ \left(\dfrac{1-t}{t}\right)  \sum \limits_{j=1}^{n}  t^j \Gamma\left(\alpha+1,\overline{E}/r^j\right)  + t^n \Gamma\left(\alpha+1,\overline{E}/r^n\right)  \right] \right\rvert_{\overline{E}_{i}}^{\overline{E}_{d}},
        \end{eqnarray}
        where $\Gamma\left(\alpha+1,y\right) = - \int e^{-y} y^{\alpha} dy $ is the incomplete gamma function.
        
        Here, we set $\overline{E}_{\hbox{leb}}$ equal to the $\overline{E}_i$ that corresponds to the first local minimum (lowest energy) of $F(\overline{E}_i)$ (Equation\ \ref{eq:diffcummul} with $F^{(p)}$ and $F^{(f)}$ from Equations\ \ref{eq:cummul1} and \ref{eq:cummul2}, respectively).
    
    \item It is straightforward to show that in this case Equation \ref{eq:logdiffcummul} can be expressed as 
     \begin{eqnarray}
             \label{eq:logdiffcummulanalyt}
             logF(x_{d}) & = & \left|\int_{x_{\hbox{leb}}}^{x_{d}}  \log\left[g_f^{(n)}\right] dx - \mathcal{F}_0 \left(x_{d} - x_{\hbox{leb}}\right) \right|.
     \end{eqnarray}
    Here, we set $\overline{E}_{\hbox{heb}}$ equal to $\overline{E}_d = \log(x_d)$ with $x_d$ equal to the last minimum (highest energy) of $logF(x_{d})$.
\end{enumerate}

Even though more consistent with the method of \S \ref{sec:fitmethod}, the above method is considerably computationally more expensive than estimating energy breaks from the approximative Equations\ \ref{eq:dist_anal_max_6} and \ref{eq:dist_anal_max_9}.


\begin{thebibliography}{}
\expandafter\ifx\csname natexlab\endcsname\relax\def\natexlab#1{#1}\fi
\providecommand{\url}[1]{\href{#1}{#1}}
\providecommand{\dodoi}[1]{doi:~\href{http://doi.org/#1}{\nolinkurl{#1}}}
\providecommand{\doeprint}[1]{\href{http://ascl.net/#1}{\nolinkurl{http://ascl.net/#1}}}
\providecommand{\doarXiv}[1]{\href{https://arxiv.org/abs/#1}{\nolinkurl{https://arxiv.org/abs/#1}}}

\bibitem[{{Adhikari} {et~al.}(2019){Adhikari}, {Khabarova}, {Zank}, \&
  {Zhao}}]{Adhikari_2019}
{Adhikari}, L., {Khabarova}, O., {Zank}, G.~P., \& {Zhao}, L.~L. 2019, \apj,
  873, 72, \dodoi{10.3847/1538-4357/ab05c6}

\bibitem[{{Alaoui} \& {Holman}(2017)}]{Alaoui_2017}
{Alaoui}, M., \& {Holman}, G.~D. 2017, \apj, 851, 78.
\newblock \doarXiv{1706.03897}

\bibitem[{Alaoui {et~al.}(2019)Alaoui, Krucker, \& Saint-Hilaire}]{Alaoui_2019}
Alaoui, M., Krucker, S., \& Saint-Hilaire, P. 2019, Solar Physics, 294, 105,
  \dodoi{10.1007/s11207-019-1495-6}

\bibitem[{Allred {et~al.}(2020)Allred, Alaoui, Kowalski, \& Kerr}]{Allred_2020}
Allred, J.~C., Alaoui, M., Kowalski, A.~F., \& Kerr, G.~S. 2020, The
  Astrophysical Journal, 902, 16, \dodoi{10.3847/1538-4357/abb239}

\bibitem[{{Antiochos}(1998)}]{Antiochos_1998}
{Antiochos}, S.~K. 1998, \apjl, 502, L181

\bibitem[{{Antiochos} {et~al.}(1999){Antiochos}, {DeVore}, \&
  {Klimchuk}}]{Antiochos_1999_I}
{Antiochos}, S.~K., {DeVore}, C.~R., \& {Klimchuk}, J.~A. 1999, \apj, 510, 485

\bibitem[{Arnold {et~al.}(2019)Arnold, Drake, Swisdak, \& Dahlin}]{Arnold_2019}
Arnold, H., Drake, J.~F., Swisdak, M., \& Dahlin, J. 2019, Physics of Plasmas,
  26, 102903, \dodoi{10.1063/1.5120373}

\bibitem[{{Arnold} {et~al.}(2021){Arnold}, {Drake}, {Swisdak}, {Guo}, {Dahlin},
  {Chen}, {Fleishman}, {Glesener}, {Kontar}, {Phan}, \& {Shen}}]{Arnold_2021}
{Arnold}, H., {Drake}, J.~F., {Swisdak}, M., {et~al.} 2021, \prl, 126, 135101,
  \dodoi{10.1103/PhysRevLett.126.135101}

\bibitem[{{Ball} {et~al.}(2018){Ball}, {Sironi}, \& {{\"O}zel}}]{Ball_2018}
{Ball}, D., {Sironi}, L., \& {{\"O}zel}, F. 2018, \apj, 862, 80.
\newblock \doarXiv{1803.05556}

\bibitem[{{B{\'a}rta} {et~al.}(2008){B{\'a}rta}, {Karlick{\'y}}, \& {{\v
  Z}emli{\v c}ka}}]{Barta_2008}
{B{\'a}rta}, M., {Karlick{\'y}}, M., \& {{\v Z}emli{\v c}ka}, R. 2008,
  \solphys, 253, 173

\bibitem[{{Battaglia} {et~al.}(2019){Battaglia}, {Kontar}, \&
  {Motorina}}]{Battaglia_2019}
{Battaglia}, M., {Kontar}, E.~P., \& {Motorina}, G. 2019, \apj, 872, 204.
\newblock \doarXiv{1901.07767}

\bibitem[{{Borovikov} {et~al.}(2017){Borovikov}, {Tenishev}, {Gombosi},
  {Guidoni}, {DeVore}, {Karpen}, \& {Antiochos}}]{Borovikov_2017}
{Borovikov}, D., {Tenishev}, V., {Gombosi}, T.~I., {et~al.} 2017, \apj, 835, 48

\bibitem[{{Brown}(1971)}]{Brown_1971}
{Brown}, J.~C. 1971, \solphys, 18, 489

\bibitem[{{Carmichael}(1964)}]{Carmichael_1964}
{Carmichael}, H. 1964, NASA Special Publication, 50, 451

\bibitem[{{Cassak} \& {Drake}(2013)}]{Cassak_2013}
{Cassak}, P.~A., \& {Drake}, J.~F. 2013, Physics of Plasmas, 20, 061207

\bibitem[{{Christe} {et~al.}(2008){Christe}, {Hannah}, {Krucker}, {McTiernan},
  \& {Lin}}]{Christe_2008}
{Christe}, S., {Hannah}, I.~G., {Krucker}, S., {McTiernan}, J., \& {Lin}, R.~P.
  2008, \apj, 677, 1385, \dodoi{10.1086/529011}

\bibitem[{Clauset {et~al.}(2009)Clauset, Shalizi, \& Newman}]{Clauset_2009}
Clauset, A., Shalizi, C.~R., \& Newman, M. E.~J. 2009, SIAM Review, 51, 661,
  \dodoi{10.1137/070710111}

\bibitem[{{Dahlin} {et~al.}(2016){Dahlin}, {Drake}, \& {Swisdak}}]{Dahlin_2016}
{Dahlin}, J.~T., {Drake}, J.~F., \& {Swisdak}, M. 2016, Physics of Plasmas, 23,
  120704.
\newblock \doarXiv{1607.03857}

\bibitem[{{Dahlin} {et~al.}(2017){Dahlin}, {Drake}, \& {Swisdak}}]{Dahlin_2017}
---. 2017, Physics of Plasmas, 24, 092110.
\newblock \doarXiv{1706.00481}

\bibitem[{{Daughton} {et~al.}(2014){Daughton}, {Nakamura}, {Karimabadi},
  {Roytershteyn}, \& {Loring}}]{Daughton_2014}
{Daughton}, W., {Nakamura}, T.~K.~M., {Karimabadi}, H., {Roytershteyn}, V., \&
  {Loring}, B. 2014, Physics of Plasmas, 21, 052307

\bibitem[{{Daughton} {et~al.}(2006){Daughton}, {Scudder}, \&
  {Karimabadi}}]{Daughton_2006}
{Daughton}, W., {Scudder}, J., \& {Karimabadi}, H. 2006, Physics of Plasmas,
  13, 072101

\bibitem[{{Dennis}(1985)}]{Dennis_1985}
{Dennis}, B.~R. 1985, \solphys, 100, 465, \dodoi{10.1007/BF00158441}

\bibitem[{{DeVore} \& {Antiochos}(2008)}]{DeVore_2008}
{DeVore}, C.~R., \& {Antiochos}, S.~K. 2008, \apj, 680, 740

\bibitem[{{Drake} {et~al.}(2019){Drake}, {Arnold}, {Swisdak}, \&
  {Dahlin}}]{Drake_2019}
{Drake}, J.~F., {Arnold}, H., {Swisdak}, M., \& {Dahlin}, J.~T. 2019, Physics
  of Plasmas, 26, 012901.
\newblock \doarXiv{1809.04568}

\bibitem[{{Drake} {et~al.}(2010){Drake}, {Opher}, {Swisdak}, \&
  {Chamoun}}]{Drake_2010}
{Drake}, J.~F., {Opher}, M., {Swisdak}, M., \& {Chamoun}, J.~N. 2010, \apj,
  709, 963.
\newblock \doarXiv{0911.3098}

\bibitem[{{Drake} {et~al.}(2005){Drake}, {Shay}, {Thongthai}, \&
  {Swisdak}}]{Drake_2005}
{Drake}, J.~F., {Shay}, M.~A., {Thongthai}, W., \& {Swisdak}, M. 2005, Physical
  Review Letters, 94, 095001

\bibitem[{{Drake} {et~al.}(2006{\natexlab{a}}){Drake}, {Swisdak}, {Che}, \&
  {Shay}}]{Drake_2006}
{Drake}, J.~F., {Swisdak}, M., {Che}, H., \& {Shay}, M.~A. 2006{\natexlab{a}},
  \nat, 443, 553

\bibitem[{{Drake} {et~al.}(2013){Drake}, {Swisdak}, \& {Fermo}}]{Drake_2013}
{Drake}, J.~F., {Swisdak}, M., \& {Fermo}, R. 2013, \apjl, 763, L5.
\newblock \doarXiv{1210.4830}

\bibitem[{{Drake} {et~al.}(2006{\natexlab{b}}){Drake}, {Swisdak}, {Schoeffler},
  {Rogers}, \& {Kobayashi}}]{Drake_2006_I}
{Drake}, J.~F., {Swisdak}, M., {Schoeffler}, K.~M., {Rogers}, B.~N., \&
  {Kobayashi}, S. 2006{\natexlab{b}}, \grl, 33, 13105

\bibitem[{Emslie(2003)}]{Emslie_2003}
Emslie, A.~G. 2003, The Astrophysical Journal, 595, L119,
  \dodoi{10.1086/378168}

\bibitem[{{Fermo} {et~al.}(2010){Fermo}, {Drake}, \& {Swisdak}}]{Fermo_2010}
{Fermo}, R.~L., {Drake}, J.~F., \& {Swisdak}, M. 2010, Physics of Plasmas, 17,
  010702.
\newblock \doarXiv{0910.4971}

\bibitem[{{Galloway} {et~al.}(2005){Galloway}, {MacKinnon}, {Kontar}, \&
  {Heland er}}]{Galloway_2005}
{Galloway}, R.~K., {MacKinnon}, A.~L., {Kontar}, E.~P., \& {Heland er}, P.
  2005, \aap, 438, 1107, \dodoi{10.1051/0004-6361:20042137}

\bibitem[{{Guidoni} {et~al.}(2016){Guidoni}, {DeVore}, {Karpen}, \&
  {Lynch}}]{Guidoni_2016}
{Guidoni}, S.~E., {DeVore}, C.~R., {Karpen}, J.~T., \& {Lynch}, B.~J. 2016,
  \apj, 820, 60.
\newblock \doarXiv{1603.01309}

\bibitem[{{Guo} {et~al.}(2015){Guo}, {Liu}, {Daughton}, \& {Li}}]{Guo_2015}
{Guo}, F., {Liu}, Y.-H., {Daughton}, W., \& {Li}, H. 2015, \apj, 806, 167.
\newblock \doarXiv{1504.02193}

\bibitem[{{Hannah} {et~al.}(2008){Hannah}, {Christe}, {Krucker}, {Hurford},
  {Hudson}, \& {Lin}}]{Hannah_2008}
{Hannah}, I.~G., {Christe}, S., {Krucker}, S., {et~al.} 2008, \apj, 677, 704,
  \dodoi{10.1086/529012}

\bibitem[{{Hayes} {et~al.}(2019){Hayes}, {Gallagher}, {Dennis}, {Ireland},
  {Inglis}, \& {Morosan}}]{Hayes_2019}
{Hayes}, L.~A., {Gallagher}, P.~T., {Dennis}, B.~R., {et~al.} 2019, \apj, 875,
  33, \dodoi{10.3847/1538-4357/ab0ca3}

\bibitem[{{Hayes} {et~al.}(2016){Hayes}, {Gallagher}, {Dennis}, {Ireland},
  {Inglis}, \& {Ryan}}]{Hayes_2016}
---. 2016, \apjl, 827, L30.
\newblock \doarXiv{1607.06957}

\bibitem[{{Hirayama}(1974)}]{Hirayama_1974}
{Hirayama}, T. 1974, \solphys, 34, 323

\bibitem[{{Holman}(2003)}]{Holman_2003_I}
{Holman}, G.~D. 2003, \apj, 586, 606

\bibitem[{{Holman} {et~al.}(2003){Holman}, {Sui}, {Schwartz}, \&
  {Emslie}}]{Holman_2003}
{Holman}, G.~D., {Sui}, L., {Schwartz}, R.~A., \& {Emslie}, A.~G. 2003, \apjl,
  595, L97

\bibitem[{{Huang} \& {Bhattacharjee}(2012)}]{Huang_2012}
{Huang}, Y.-M., \& {Bhattacharjee}, A. 2012, Physical Review Letters, 109,
  265002.
\newblock \doarXiv{1211.6708}

\bibitem[{{Hudson}(1972)}]{Hudson_1972}
{Hudson}, H.~S. 1972, \solphys, 24, 414

\bibitem[{{Inglis} \& {Dennis}(2012)}]{Inglis_2012}
{Inglis}, A.~R., \& {Dennis}, B.~R. 2012, \apj, 748, 139.
\newblock \doarXiv{1303.6309}

\bibitem[{{Inglis} \& {Gilbert}(2013)}]{Inglis_2013}
{Inglis}, A.~R., \& {Gilbert}, H.~R. 2013, \apj, 777, 30.
\newblock \doarXiv{1307.2874}

\bibitem[{{Inglis} {et~al.}(2016){Inglis}, {Ireland}, {Dennis}, {Hayes}, \&
  {Gallagher}}]{Inglis_2016}
{Inglis}, A.~R., {Ireland}, J., {Dennis}, B.~R., {Hayes}, L., \& {Gallagher},
  P. 2016, \apj, 833, 284, \dodoi{10.3847/1538-4357/833/2/284}

\bibitem[{{Karlick{\'y}}(2004)}]{Karlicky_2004}
{Karlick{\'y}}, M. 2004, \aap, 417, 325

\bibitem[{{Karlick{\'y}} \& {B{\'a}rta}(2007)}]{Karlicky_2007}
{Karlick{\'y}}, M., \& {B{\'a}rta}, M. 2007, \aap, 464, 735

\bibitem[{{Karpen} {et~al.}(2012){Karpen}, {Antiochos}, \&
  {DeVore}}]{Karpen_2012}
{Karpen}, J.~T., {Antiochos}, S.~K., \& {DeVore}, C.~R. 2012, \apj, 760, 81

\bibitem[{{Kliem} {et~al.}(2000){Kliem}, {Karlick{\'y}}, \&
  {Benz}}]{Kliem_2000}
{Kliem}, B., {Karlick{\'y}}, M., \& {Benz}, A.~O. 2000, \aap, 360, 715

\bibitem[{{Kontar} {et~al.}(2008){Kontar}, {Dickson}, \& {Ka{\v
  s}parov{\'a}}}]{Kontar_2008}
{Kontar}, E.~P., {Dickson}, E., \& {Ka{\v s}parov{\'a}}, J. 2008, \solphys,
  252, 139.
\newblock \doarXiv{0805.1470}

\bibitem[{{Kontar} {et~al.}(2015){Kontar}, {Jeffrey}, {Emslie}, \&
  {Bian}}]{Kontar_2015}
{Kontar}, E.~P., {Jeffrey}, N. L.~S., {Emslie}, A.~G., \& {Bian}, N.~H. 2015,
  \apj, 809, 35, \dodoi{10.1088/0004-637X/809/1/35}

\bibitem[{{Kopp} \& {Pneuman}(1976)}]{Kopp_1976}
{Kopp}, R.~A., \& {Pneuman}, G.~W. 1976, \solphys, 50, 85

\bibitem[{{Krucker} {et~al.}(2010){Krucker}, {Hudson}, {Glesener}, {White},
  {Masuda}, {Wuelser}, \& {Lin}}]{Krucker_2010}
{Krucker}, S., {Hudson}, H.~S., {Glesener}, L., {et~al.} 2010, \apj, 714, 1108

\bibitem[{{Krucker} \& {Lin}(2008)}]{Krucker_2008}
{Krucker}, S., \& {Lin}, R.~P. 2008, \apj, 673, 1181

\bibitem[{{Krucker} {et~al.}(2008){Krucker}, {Battaglia}, {Cargill},
  {Fletcher}, {Hudson}, {MacKinnon}, {Masuda}, {Sui}, {Tomczak}, {Veronig},
  {Vlahos}, \& {White}}]{Krucker_2008_rev}
{Krucker}, S., {Battaglia}, M., {Cargill}, P.~J., {et~al.} 2008, \aapr, 16, 155

\bibitem[{{Kumar} \& {Cho}(2013)}]{Kumar_2013}
{Kumar}, P., \& {Cho}, K.-S. 2013, \aap, 557, A115.
\newblock \doarXiv{1307.3910}

\bibitem[{{Kumar} \& {Innes}(2013)}]{Kumar_2013_I}
{Kumar}, P., \& {Innes}, D.~E. 2013, \solphys, 288, 255.
\newblock \doarXiv{1307.3720}

\bibitem[{{le Roux} {et~al.}(2018){le Roux}, {Zank}, \&
  {Khabarova}}]{le_Roux_2018}
{le Roux}, J.~A., {Zank}, G.~P., \& {Khabarova}, O.~V. 2018, \apj, 864, 158

\bibitem[{{le Roux} {et~al.}(2015){le Roux}, {Zank}, {Webb}, \&
  {Khabarova}}]{le_Roux_2015}
{le Roux}, J.~A., {Zank}, G.~P., {Webb}, G.~M., \& {Khabarova}, O. 2015, \apj,
  801, 112, \dodoi{10.1088/0004-637X/801/2/112}

\bibitem[{{Li} {et~al.}(2018){Li}, {Guo}, {Li}, \& {Li}}]{Li_2018}
{Li}, X., {Guo}, F., {Li}, H., \& {Li}, S. 2018, \apj, 866, 4.
\newblock \doarXiv{1807.03427}

\bibitem[{{Li} {et~al.}(2019){Li}, {Guo}, {Li}, {Stanier}, \&
  {Kilian}}]{Li_2019}
{Li}, X., {Guo}, F., {Li}, H., {Stanier}, A., \& {Kilian}, P. 2019, \apj, 884,
  118, \dodoi{10.3847/1538-4357/ab4268}

\bibitem[{{Liu} {et~al.}(2013){Liu}, {Chen}, \& {Petrosian}}]{Liu_2013}
{Liu}, W., {Chen}, Q., \& {Petrosian}, V. 2013, \apj, 767, 168.
\newblock \doarXiv{1303.3321}

\bibitem[{{Loureiro} {et~al.}(2007){Loureiro}, {Schekochihin}, \&
  {Cowley}}]{Loureiro_2007}
{Loureiro}, N.~F., {Schekochihin}, A.~A., \& {Cowley}, S.~C. 2007, Physics of
  Plasmas, 14, 100703

\bibitem[{{Masuda} {et~al.}(1994){Masuda}, {Kosugi}, {Hara}, {Tsuneta}, \&
  {Ogawara}}]{Masuda_1994}
{Masuda}, S., {Kosugi}, T., {Hara}, H., {Tsuneta}, S., \& {Ogawara}, Y. 1994,
  \nat, 371, 495

\bibitem[{{McTiernan} {et~al.}(2019){McTiernan}, {Caspi}, \&
  {Warren}}]{McTiernan_2019}
{McTiernan}, J.~M., {Caspi}, A., \& {Warren}, H.~P. 2019, \apj, 881, 161.
\newblock \doarXiv{1805.12285}

\bibitem[{{Mei} {et~al.}(2012){Mei}, {Shen}, {Wu}, {Lin}, {Murphy}, \&
  {Roussev}}]{Mei_2012}
{Mei}, Z., {Shen}, C., {Wu}, N., {et~al.} 2012, \mnras, 425, 2824

\bibitem[{{Oka} {et~al.}(2018){Oka}, {Birn}, {Battaglia}, {Chaston}, {Hatch},
  {Livadiotis}, {Imada}, {Miyoshi}, {Kuhar}, {Effenberger}, {Eriksson},
  {Khotyaintsev}, \& {Retin{\`o}}}]{Oka_2018}
{Oka}, M., {Birn}, J., {Battaglia}, M., {et~al.} 2018, \ssr, 214, 82.
\newblock \doarXiv{1805.09278}

\bibitem[{{Petrosian} {et~al.}(2002){Petrosian}, {Donaghy}, \&
  {McTiernan}}]{Petrosian_2002}
{Petrosian}, V., {Donaghy}, T.~Q., \& {McTiernan}, J.~M. 2002, \apj, 569, 459

\bibitem[{{Saint-Hilaire} \& {Benz}(2005)}]{Saint_Hilaire_2005}
{Saint-Hilaire}, P., \& {Benz}, A.~O. 2005, A\&A, 435, 743

\bibitem[{Samtaney {et~al.}(2009)Samtaney, Loureiro, Uzdensky, Schekochihin, \&
  Cowley}]{Samtaney_2009}
Samtaney, R., Loureiro, N.~F., Uzdensky, D.~A., Schekochihin, A.~A., \& Cowley,
  S.~C. 2009, Phys. Rev. Lett., 103, 105004

\bibitem[{{Shen} {et~al.}(2013){Shen}, {Lin}, {Murphy}, \&
  {Raymond}}]{Shen_2013}
{Shen}, C., {Lin}, J., {Murphy}, N.~A., \& {Raymond}, J.~C. 2013, Physics of
  Plasmas, 20, 072114

\bibitem[{{Sturrock}(1966)}]{Sturrock_1966}
{Sturrock}, P.~A. 1966, \nat, 211, 695

\bibitem[{{Su} {et~al.}(2011){Su}, {Holman}, \& {Dennis}}]{Su_2011}
{Su}, Y., {Holman}, G.~D., \& {Dennis}, B.~R. 2011, \apj, 731, 106,
  \dodoi{10.1088/0004-637X/731/2/106}

\bibitem[{{Takasao} {et~al.}(2016){Takasao}, {Asai}, {Isobe}, \&
  {Shibata}}]{Takasao_2016}
{Takasao}, S., {Asai}, A., {Isobe}, H., \& {Shibata}, K. 2016, \apj, 828, 103.
\newblock \doarXiv{1611.00108}

\bibitem[{{Tandberg-Hanssen} \& {Emslie}(1988)}]{Tandberg_Hanssen_1988}
{Tandberg-Hanssen}, E., \& {Emslie}, A.~G. 1988, {The physics of solar flares}
  (Cambridge: University Press)

\bibitem[{Uzdensky {et~al.}(2010)Uzdensky, Loureiro, \&
  Schekochihin}]{Uzdensky_2010}
Uzdensky, D.~A., Loureiro, N.~F., \& Schekochihin, A.~A. 2010, Phys. Rev.
  Lett., 105, 235002

\bibitem[{Virkar \& Clauset(2014)}]{Virkar_2014}
Virkar, Y., \& Clauset, A. 2014, Ann. Appl. Stat., 8, 89,
  \dodoi{10.1214/13-AOAS710}

\bibitem[{{Zank} {et~al.}(2014){Zank}, {le Roux}, {Webb}, {Dosch}, \&
  {Khabarova}}]{Zank_2014}
{Zank}, G.~P., {le Roux}, J.~A., {Webb}, G.~M., {Dosch}, A., \& {Khabarova}, O.
  2014, \apj, 797, 28

\bibitem[{{Zhao} {et~al.}(2018){Zhao}, {Zank}, {Khabarova}, {Du}, {Chen},
  {Adhikari}, \& {Hu}}]{Zhao_2018}
{Zhao}, L.~L., {Zank}, G.~P., {Khabarova}, O., {et~al.} 2018, \apjl, 864, L34,
  \dodoi{10.3847/2041-8213/aaddf6}

\bibitem[{{Zhao} {et~al.}(2019){Zhao}, {Xia}, {Van Doorsselaere}, {Keppens}, \&
  {Gan}}]{Zhao_2019}
{Zhao}, X., {Xia}, C., {Van Doorsselaere}, T., {Keppens}, R., \& {Gan}, W.
  2019, \apj, 872, 190, \dodoi{10.3847/1538-4357/ab0284}

\bibitem[{{Zharkova} {et~al.}(2011){Zharkova}, {Arzner}, {Benz}, {Browning},
  {Dauphin}, {Emslie}, {Fletcher}, {Kontar}, {Mann}, {Onofri}, {Petrosian},
  {Turkmani}, {Vilmer}, \& {Vlahos}}]{Zharkova_2011}
{Zharkova}, V.~V., {Arzner}, K., {Benz}, A.~O., {et~al.} 2011, \ssr, 159, 357.
\newblock \doarXiv{1110.2359}

\end{thebibliography}
\end{document}